\documentclass[a4paper,11pt]{article}
\pdfoutput=1 

\usepackage{jheppub} 
\usepackage{mathrsfs}
\usepackage{amsfonts}
\usepackage{amsmath,amssymb,amsfonts}
\usepackage{slashed}
\usepackage{array}
\usepackage{float}
\usepackage{verbatim}
\usepackage{epsfig}
\usepackage{graphicx}
\usepackage{color}
\usepackage[colorlinks=true, linkcolor=blue, citecolor=blue, urlcolor=blue]{hyperref}
\usepackage[dvipsnames]{xcolor}
\usepackage{multirow}
\usepackage{lipsum}
\usepackage{cleveref}
\usepackage{comment}
\usepackage{physics}
\usepackage{mathtools}
\hypersetup{
	colorlinks=true,
	linkcolor=blue,
	filecolor=magenta,      
	urlcolor=blue,}

\usepackage[T1]{fontenc} 

\title{Lattice QCD calculation of {the pion distribution amplitude} with domain wall fermions at physical pion mass}

\author[a]{Ethan Baker}
\author[d]{Dennis Bollweg}
\author[b]{Peter Boyle}
\author[c]{Ian Cloët}
\author[c]{Xiang Gao}
\author[b]{Swagato Mukherjee}
\author[b]{Peter Petreczky}
\author[c]{Rui Zhang}
\author[c]{Yong Zhao}
\affiliation[a]{Department of Physics and Astronomy, Haverford College, Haverford, PA 19041, USA}
\affiliation[b]{Physics Department, Brookhaven National Laboratory, Bldg. 510A, Upton, New York 11973, USA}
\affiliation[c]{Physics Division, Argonne National Laboratory, Lemont, IL 60439, USA}
\affiliation[d]{Computational Science Initiative, Brookhaven National Laboratory, Bldg. 725, Upton, New York 11973, USA}

\abstract{
	We present a direct lattice QCD calculation of the $x$-dependence of the pion distribution amplitude (DA), which is performed using the quasi-DA in large momentum effective theory on a domain-wall fermion ensemble at physical quark masses and spacing $a\approx 0.084$ fm. The bare quais-DA matrix elements are renormalized in the hybrid scheme and matched to $\overline{\rm MS}$ with a subtraction of the leading renormalon in the Wilson-line mass. For the first time, we include threshold resummation in the perturbative matching onto the light-cone DA, which resums the large logarithms in the soft gluon limit at next-to-next-to-leading log. The resummed results show controlled scale-variation uncertainty within the range of momentum fraction $x\in[0.25,0.75]$ at the largest pion momentum $P_z\approx 1.85$~GeV. In addition, we apply the same analysis to quasi-DAs from a highly-improved-staggered-quark ensemble at physical pion mass and $a=0.076$ fm. By comparison we find with $2\sigma$ confidence level that the DA obtained from chiral fermions is flatter and lower near $x=0.5$.
}

\begin{document}

\maketitle
\flushbottom

\section{Introduction}
Understanding the structure of pions is of special importance as they are the lightest pseudo-Nambu-Goldstone bosons of chiral symmetry breaking in strong interactions. {One particular quantity of interest is the pion distribution amplitude (DA) $\phi(x)$ which describes the probability amplitude of finding pion on the light-cone in a quark-antiquark pair Fock state, each carrying a momentum fraction of $x$ and $1-x$.} The rich phenomenology of the pion DA originates from its universality as inputs to exclusive processes and form factors at large momentum transfer $Q^2\gg \Lambda_{\rm QCD}^2$. 
{For example, the scattering amplitudes of semileptonic $B$-meson decay~\cite{Beneke:1999br,Beneke:2001ev} and deeply virtual meson production processes~\cite{Collins:1996fb}, which have been used to probe physics beyond the Standard Model and to extract the generalized parton distributions, respectively, are proportional to convolutions involving the pion DA in the $x$-space.}
Therefore, knowing the $x$-dependence of DAs is key to making predictions for the hard exclusive processes. However, they are only weakly constrained by experiments so far~\cite{CLEO:1997fho,CELLO:1990klc,BaBar:2009rrj,Belle-II:2018jsg}, which makes it highly desirable to calculate their $x$-dependence from first-principle methods like lattice QCD.

The pion DA $\phi(x,\mu)$ is defined from a light-cone correlation as
\begin{align}\label{eq.light-coneDA}
{if_\pi}\phi(x,\mu)=& \int\frac{d \eta^-}{2\pi}e^{ixP^+\eta^-}\times\langle{0}|{\overline{\psi}(0)\gamma_5\gamma_+ W(0,\eta^-)\psi(\eta^-)}|{\pi(P)}\rangle,
\end{align}
where {$f_\pi$ is the pion decay constant}, $W(0,\eta^-)=\hat{\mathcal{P}}\exp\left[-ig\int^{\eta^-}_0\!\!ds\,n_{\mu}A^{\mu}(ns)\right]$ is the Wilson line between the two light-cone coordinates $0$ and $\eta^-=(\eta^0+\eta^3)/\sqrt{2}$ with $\hat{\mathcal{P}}$ denoting the path-ordering operator. {Due to the real-time dependence of light-cone correlations, } it is impossible to directly calculate the full pion DA on a Euclidean lattice.

There are different approaches to extract information of the pion DA from lattice QCD, including the calculation of its moments $\langle (2x-1)^n\rangle$~\cite{Kronfeld:1984zv,DelDebbio:2002mq,Braun:2006dg,Arthur:2010xf,Bali:2017ude,RQCD:2019osh}, short distance {factorization (SDF) of nonlocal} correlations~\cite{Braun:2007wv,Braun:2015axa,Bali:2018spj,Detmold:2005gg,Detmold:2021qln,Gao:2022vyh}, and the \textit{large momentum effective theory} (LaMET)~\cite{Ji:2013dva,Ji:2014gla,Ji:2020ect,Zhang:2017bzy,Zhang:2017zfe,Zhang:2020gaj,LatticeParton:2022zqc,Holligan:2023rex}. The calculation of moments is based on the light-cone operator product expansion (OPE) of the pion-DA correlator, which encodes the DA moments in twist-two local operators~\cite{Kronfeld:1984zv}. The bottleneck of this method is the increasing noise in higher moments and  {power-divergent operator mixings} for $n>3$. In the SDF approach, the spatial correlation function in the pion state can have an OPE~\cite{Izubuchi:2018srq} or be factorized into the convolution of the light-cone correlation and a perturbative matching kernel at a distance $|z|\ll \Lambda_{\rm QCD}^{-1}$. Therefore, this method can provide information of the first few moments, which in principle can go beyond $n=3$, {or the light-cone correlation within a limited range}. However, it still requires a model assumption of the shape of DA to reconstruct its $x$-dependence~\cite{Bali:2018spj,Gao:2022vyh}. 
{The LaMET approach provides a direct calculation of the $x$-dependence through an effective theory expansion in the momentum space, up to power corrections suppressed by the parton momenta $xP_z$ and $(1-x)P_z$}, which allows for a reliable calculation of the DA within a moderate range of $x$. 

Since the first LaMET calculation of the pion DA~\cite{Zhang:2017bzy}, various lattice artifacts and theoretical systematics have been studied. 
{One of the key systematics is the lattice renormalization of the quasi-DA correlators, which suffer from the linear power divergence that must be subtracted at all distances~\cite{Chen:2016fxx,Ji:2017oey,Green:2017xeu,Ishikawa:2017faj}. The regularization-independent momentum subtraction (RI/MOM)~\cite{Constantinou:2017sej,Alexandrou_2017,Chen:2017mzz,Stewart:2017tvs} and ratio~\cite{Orginos:2017kos,Fan:2020nzz} schemes were proposed to cancel the linear and other ultraviolet (UV) divergences, but both introduce uncontrolled non-perturbative effects at large distance for LaMET calculation. Then, the hybrid scheme~\cite{Ji:2020brr} was proposed to overcome this problem with a subtraction of Wilson line mass at long distances~\cite{LatticePartonCollaborationLPC:2021xdx,Gao:2021dbh}, but there is still a remaining linear renormalon ambiguity of order $\Lambda_{\rm QCD}$ that contributes to a linear power correction in the LaMET expansion. Recently, this issue was resolved with the leading-renormalon resummation (LRR) method~\cite{Holligan:2023rex,Zhang:2023bxs}, which removes such an ambiguity in lattice renormalization and LaMET matching, thus improving the power accuracy to sub-leading order. Apart from lattice renormalization, there has also been significant progress in the perturbative matching, from the next-to-leading order (NLO) matching kernel for quasi-DA~\cite{Zhang:2017bzy,Zhang:2017zfe,Bali:2017gfr,Xu:2018mpf,Liu:2018tox} to the renormalization group resummation (RGR)~\cite{Gao:2021hxl,Su:2022fiu} and threshold resummation~\cite{Gao:2021hxl,Ji:2023pba,Liu:2023onm}.
On the numerical side, the first continuum extrapolation of the DA was carried out in Ref.~\cite{Zhang:2020gaj} with the RI/MOM scheme at unphysical pion masses, which observed a sensitive dependence on the pion mass. This calculation was succeeded by a continuum extrapolation at physical pion mass with NLO hybrid-scheme renormalization and matching, but without LRR~\cite{LatticeParton:2022zqc}. The LRR method was first implemented in Ref.~\cite{Holligan:2023rex} with NLO matching and RGR. Meanwhile, the pion DA moments have also been calculated using the SDF approach at NLO accuracy~\cite{Bali:2018spj,Gao:2022vyh}.
Nevertheless, no lattice calculation in the literature has included the threshold resummation yet.
}

So far, all the lattice calculations of pion DA $x$-dependence have been performed with highly-improved-staggered-quark (HISQ) fermion actions~\cite{Kogut:1974ag,Follana:2006rc} or clover fermion actions~\cite{Curci:1983an}, both of which explicitly break the chiral symmetry. Since the pions are the pseudo-Nambu-Goldstone bosons of chiral symmetry breaking in QCD, it is then very interesting to study how the chiral symmetry of lattice action affects their structure such as the DA. The overlap fermion~\cite{Narayanan:1993ss,Narayanan:1993sk,Narayanan:1993zzh,Narayanan:1994gw} and domain wall fermion (DWF) actions~\cite{Kaplan:1992bt,Shamir:1993zy,Furman:1994ky} are known to preserve chiral symmetry on the lattice, although these ensembles are more expensive to generate. Therefore, performing the same calculation on chiral fermion ensembles will provide us with knowledge about the effects of chiral symmetry breaking on the pion structure. 

In this work, we present the first direct calculation of the $x$-dependence of pion DA on a lattice ensemble with the DWF fermion action at physical pion mass and spacing $a\approx 0.084$ fm. 
{We renormalize the bare pion quasi-DA matrix elements with the hybrid scheme, and then match them to the continuum $\overline{\rm MS}$ scheme with LRR. 
After Fourier transforming to the $x$-space, we match the quasi-DA to the light-cone with next-to-next-to-leading logarithmic (NNLL) threshold resummation and NLO matching. 
For the first time, we reformulate the recently developed threshold resummation technique for the quasi-PDF case~\cite{Ji:2023pba} to work for the quasi-DA, which improves the estimate of systematic uncertainties near the end-point regions $x\to0$ and $x\to1$.
With the reformulated threshold factorization, we also find out that the complicated two-scale resummation~\cite{Holligan:2023rex} of the Efremov-Radyushkin-Brodsky-Lepage (ERBL) logarithms~\cite{Efremov:1978rn,Efremov:1979qk,Lepage:1979zb,Lepage:1980fj} is now factorized into two separate pieces with each involving only a single physical scale, thus making it straightforward to {understand and implement the RGR}. Finally, we observe a notably flat DA in the region {$x\in[0.25,0.75]$}, while the results beyond this range become unreliable due to {the breakdown of perturbation theory}. Moreover, compared with the same analysis applied to the HISQ data~\cite{Gao:2022vyh} {at physical pion mass and lattice spacing $a=0.076$ fm}, we observe that the pion DA from the DWF ensemble is slightly flatter near $x=0.5$ than that from the HISQ ensemble.  Our results suggest that the pion structure is not sensitive to chiral symmetry on the lattice.

This work is organized as follows. In Sec.~\ref{sec:lat_setup}, we present our lattice setup of the calculation, and show how the raw lattice data are processed to extract the matrix elements of pion DA. In Sec.~\ref{sec:quasi_da}, we analyze the bare matrix elements to extract the DA moments using the SDF or OPE approach, and discuss how the matrix elements are properly and consistently renormalized in the hybrid scheme with LRR to obtain $x$-dependent quasi-DA. In Sec.~\ref{sec:matching}, we derive the formalism to resum both ERBL and threshold logarithms in the perturbative matching and implement the improved matching to extract the light-cone DA, and compare with the same analysis applied to data on a HISQ ensemble~\cite{Gao:2022vyh}. Finally, we conclude in Sec.~\ref{sec:conlusion}.

\section{Lattice set up}\label{sec:lat_setup}
In this calculation, we used a 2+1-flavor domain-wall gauge ensemble generated by RBC and UKQCD Collaborations of size $N_s^3\times N_t\times N_5 =64^3\times 128\times12$, denoted by 64I~\cite{RBC:2023pvn}. The quark masses are at the physical point and the lattice spacing is $a=0.0836$ fm. 
55 gauge configurations were used in this calculation. The quark propagators are evaluated from Coulomb-gauge-fixed configurations using deflation based solver with 2000 eigen vectors. In addition, we used the boosted Gaussian momentum smearing~\cite{Bali:2016lva} to improve the signal. The Gaussian radius was set to be $r_G$ = 0.58 fm. We chose the quark boost parameter $j_z$ to be 0 and 6~\cite{Gao:2021xsm, Gao:2020ito} which are optimal to hadron momentum $P_z=2\pi n_z/(N_sa)$ with $n_z$ = 0 and 8, so that the largest momentum in our calculation is $P_z=1.85$ GeV. Since only two-point functions are involved in this calculation~\cite{Gao:2022vyh}, measurements at other momenta ($n_z\in[0,3]$ for $j_z=0$ and $n_z\in[4,8]$ for $j_z=6$) were also computed through contractions using the same profiled quark propagator. To increase the statistics, we ultilzed the All Mode Averaging (AMA) technique~\cite{Shintani:2014vja} with 2 exact and 128 sloppy sources for momenta $n_z\in[4,8]$, while 1 exact and 32 sloppy sources for $n_z=[0,3]$. The tolerance of the exact and sloppy sources are $10^{-8}$ and $10^{-4}$ respectively.  To suppress the ultraviolet (UV) fluctuations and enhance the signal-to-noise ratio of the matrix elements with long Wilson links, we employed Wilson flow~\cite{Luscher:2010iy}, with a flow time $t_F=1.0$ (roughly corresponds to a smearing radius $\sqrt{8a^2}$). Utilizing the symmetry of the data, we further average the forward and backward correlators at Euclidean time slice $\tau$ and $N_t-\tau$, as well as averaging the quark bilinear separation $z$ and $-z$, with the corresponding parity. In total, we have effectively 28,160 measurements for the largest momentum $P_z=1.85$ GeV.

In accordance to the quasi-DA definition
\begin{align}
    if_\pi\tilde{\phi}_{t/z}(x,P_z)=\int\frac{dz}{2\pi}e^{izP_z(1/2-x)}
    \times\langle \pi(P)|\bar{\psi}(-\frac{z}{2})\gamma_{t/z}\gamma_5W(-\frac{z}{2},\frac{z}{2})\psi(\frac{z}{2})|0\rangle,
\end{align}
we measure three different types of correlators, $C_{\pi\pi}$, $C_{\pi O_0}$, and $C_{\pi O_3}$,  to extract the quasi-DA matrix elements:
\begin{align}
    C_{\pi\pi}(t)&=\langle O_\pi(0)O^\dagger_\pi(t)\rangle, \nonumber\\
    C_{\pi O_0}(t,z)&=\langle O_\pi(0)\bar{\psi}(-\frac{z}{2},t)\gamma_t\gamma_5W(-\frac{z}{2},\frac{z}{2})\psi(\frac{z}{2},t)\rangle,\\
    C_{\pi O_3}(t,z)&=\langle O_\pi(0)\bar{\psi}(-\frac{z}{2},t)\gamma_z\gamma_5W(-\frac{z}{2},\frac{z}{2})\psi(\frac{z}{2},t)\rangle,\nonumber
\end{align}
where $O_\pi=\bar{\psi}\gamma_5\psi$ is the smeared pseudo-scalar interpolator, that has an overlap with pion ground state $c_0=\langle O_\pi|\pi\rangle$. Here $W(-\frac{z}{2},\frac{z}{2})\equiv\Pi_{i=0}^{z-1} U_z(-\frac{z}{2}+i\hat{z},t)$ is the spatial Wilson line that keeps the non-local operator gauge invariant. 
The Gaussian momentum smearing is applied to the pion interpolator $O_\pi(x,t)$, so $C_{\pi\pi}$ is smeared at both the source and the sink, while $C_{\pi O_0}$ and $C_{\pi O_3}$ are smeared only at the source. Note that in general there could be a mixing between the non-local operators $\bar{\psi}\gamma_5\gamma_tW_z\psi$ and  $\bar{\psi}\gamma_x\gamma_y W_z\psi${~\cite{Constantinou:2017sej,Chen:2017mie}}, but such a mixing is proportional to the explicit chiral symmetry breaking, thus vanishes specifically on the DWF ensemble.

By expanding these correlators in a tower of energy eigenstates, we can extract the energy spectrum and the coefficients,
\begin{align}
\label{eq:energy_expansion}
    C_{\pi\pi}(t)&=\sum A^{\pi}_i(e^{-E_i t}+e^{-E_i (N_t-t)}),\nonumber\\
    C_{\pi O_0}(t,z)&=\sum A^{O_0}_i(z)(e^{-E_i t}+e^{-E_i (N_t-t)}),\\
    C_{\pi O_3}(t,z)&=\sum A^{O_3}_i(z)(e^{-E_i t}+e^{-E_i (N_t-t)}),\nonumber
\end{align}
with the ground-state coefficients
\begin{align}\label{eq:kinetiC_Factors}
    A^{\pi}_0&=\frac{|\langle O_\pi|\pi\rangle|^2}{2E_0},\nonumber\\
    A^{O_0}_0(z)&=\frac{\langle O_\pi|\pi\rangle}{2E_0} f_\pi H_{\gamma_t\gamma_5}(z)E_0,\\
    A^{O_3}_0(z)&=\frac{\langle O_\pi|\pi\rangle}{2E_0} if_\pi H_{\gamma_z\gamma_5}(z)P_z,\nonumber
\end{align}
where $H_{\gamma_{t/z}\gamma_5}(z)$ are the matrix elements of pion quasi-DA, which is normalized to $H_{\gamma_{t/z}\gamma_5}(0)=1$.
The $P_z=0$ correlators are non-vanishing for $C_{\pi\pi}(t)$ and $C_{\pi O_0}(t,z)$, so they could be used to extract the pion mass. 

{The effective mass for the $C_{\pi\pi}$ correlator at different momenta are shown in Fig.~\ref{fig:eff_mass}. At small Euclidean time, the excited states effect is important, thus making the effective masses of the smeared-smeared $C_{\pi\pi}$ correlator larger than the actual ground state energies.
At large Euclidean time, the effective masses decays to approach ground state plateaus, which are consistent with the dispersion relation $E(P_z)=\sqrt{P_z^2+m^2}$ from zero momentum correlators, plotted as colored lines.}
\begin{figure}[!thbp]
    \centering
    \includegraphics[width=0.45\textwidth]{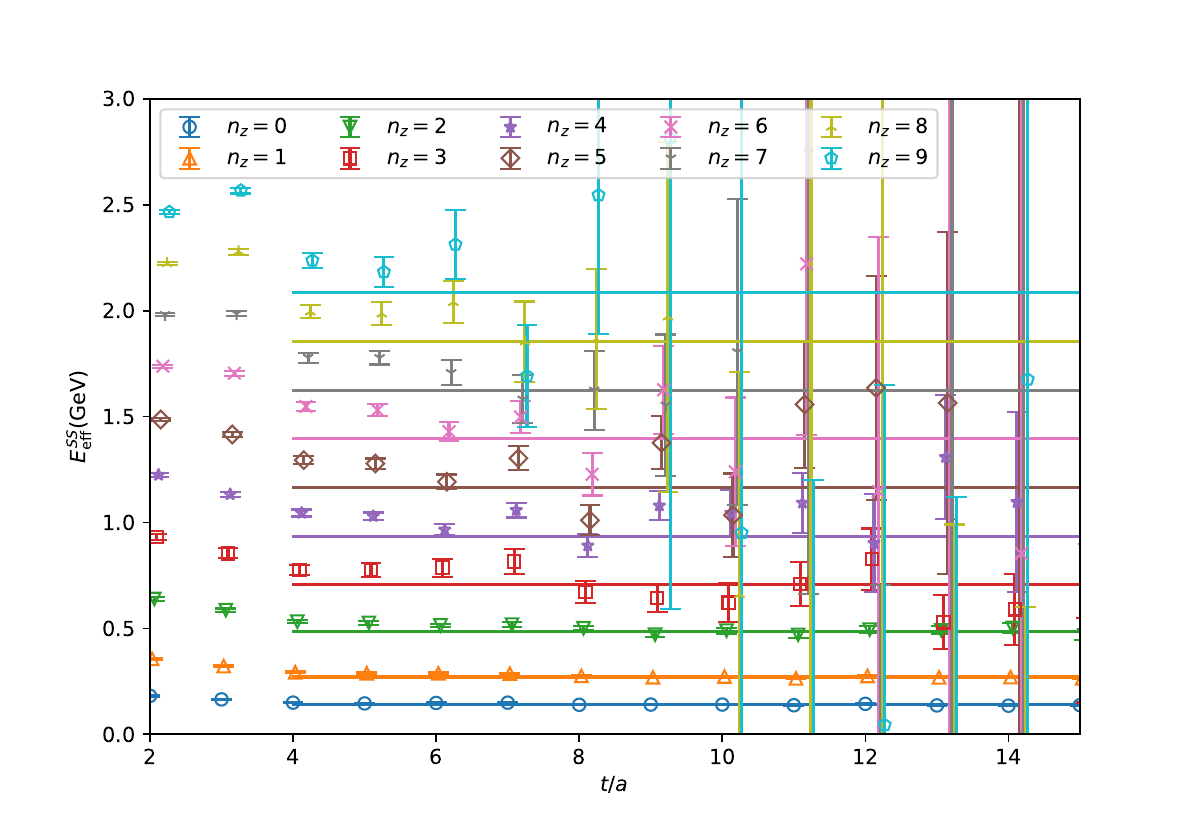}
    \caption{
    Effective mass from the $C_{\pi \pi}(t,0)$ correlators. 
    The straight lines represent the energy calculated from dispersion relation $E(P_z)=\sqrt{P_z^2+m^2}$. }
    \label{fig:eff_mass}
\end{figure}

{Thus we two-state fit the energy spectrum at $z=0$ using the dispersion relation as a prior for $E_0$. The first-excited state of the pion using smeared correlators on the lattice has been studied in previous works~\cite{Gao:2021hvs,Gao:2021xsm}. In these works 
it was suggested that the first excited state is the  $\pi(1300)$ state. 
{Therefore, the energy of the first excited pion state for different moment can be estimated using the dispersion relation
and the mass of $\pi(1300)$ state. We use this estimate as a 
prior for $E_1$, with a width of $0.5$~GeV. }
The first-excited state energies from two-state fit are shown in Fig.~\ref{fig:two_state_energy} {as function of 
$t_{min}$, with $t_{min}$ being the lower limit of the fit interval.}
By including the excited state contribution, we are able to utilize data at smaller Euclidean time, where the plateau of effective mass has not been reached, and still get good fit quality. } Two examples of the fits at $n_z=8$ and $z=0$ and $z=3a$ with different fit ranges are shown in Fig.~\ref{fig:fit_examples}. To compensate the exponential decay in Euclidean time, we multiply the correlators with a factor of $e^{\bar{E}_0 t}$. There is a good consistency between data and our $t_{\rm min}=4a$ fit bands. {The fitted ground state $E_0$ and the first excited state $E_1$ are shown in Fig.~\ref{fig:dispersion} as a function of $n_z$. These results are consistent with the priors based on the dispersion relation, depicted as blue and orange bands.}

\begin{figure}[!htbp]
    \centering
    \includegraphics[width=0.45\textwidth]{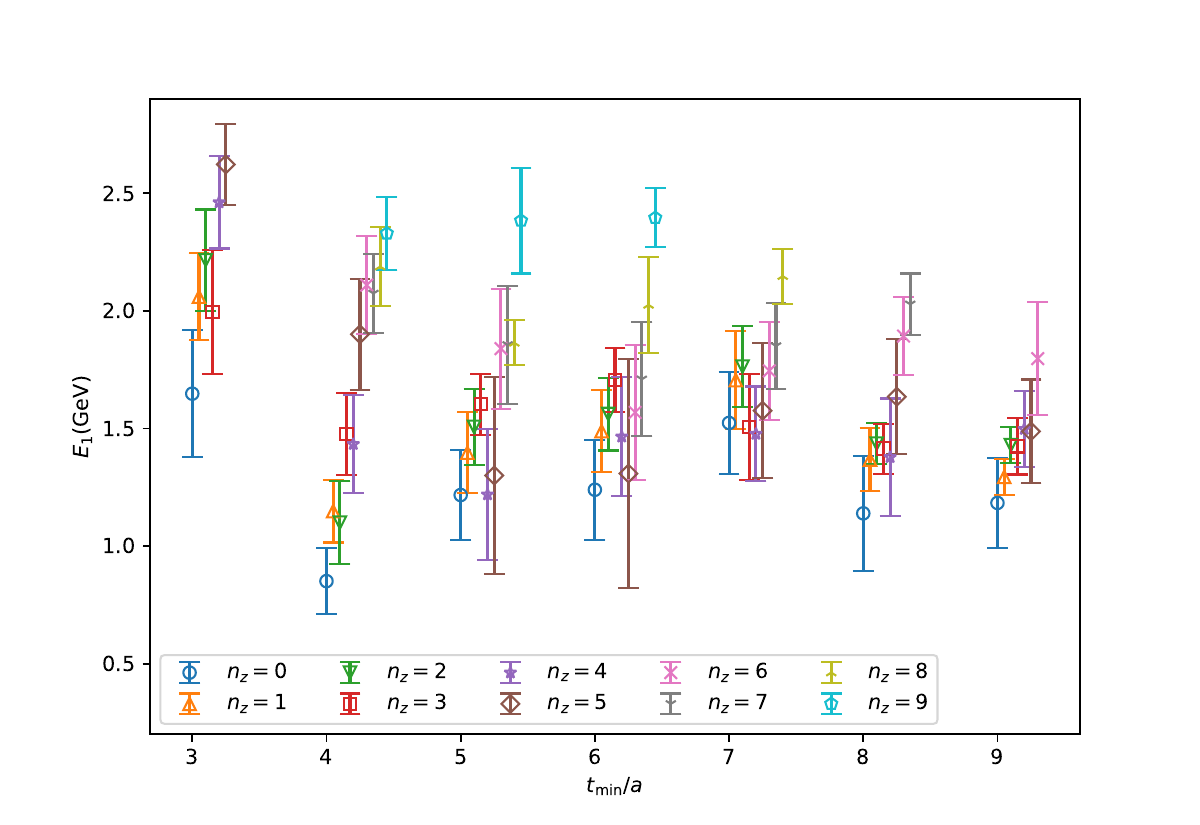}
    \caption{First excited state energy from two-state fit {as function of $t_{min}$ for different pion momenta. }}
    \label{fig:two_state_energy}
\end{figure}

\begin{figure}[!htbp]
    \centering
    \includegraphics[width=0.45\textwidth]{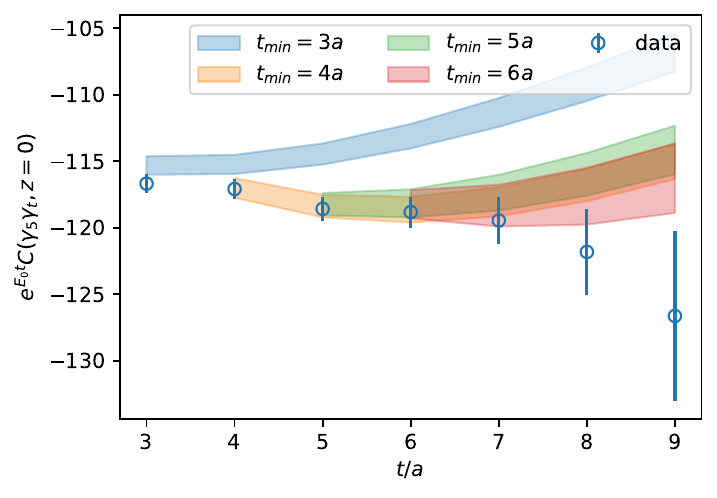}
    \includegraphics[width=0.45\textwidth]{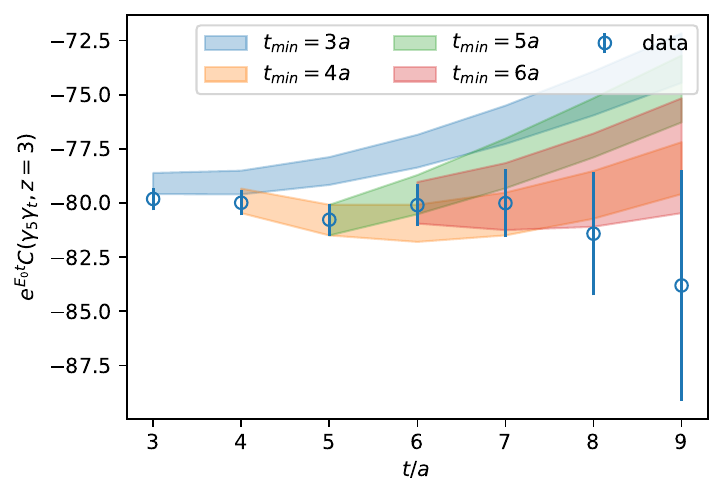}
    \caption{Examples of fitted correlators at $z=0$ (left) and $z=3a$ (right) with different $t_{\rm min}$. }
    \label{fig:fit_examples}
\end{figure}

\begin{figure}[!htbp]
    \centering
    \includegraphics[width=0.45\textwidth]{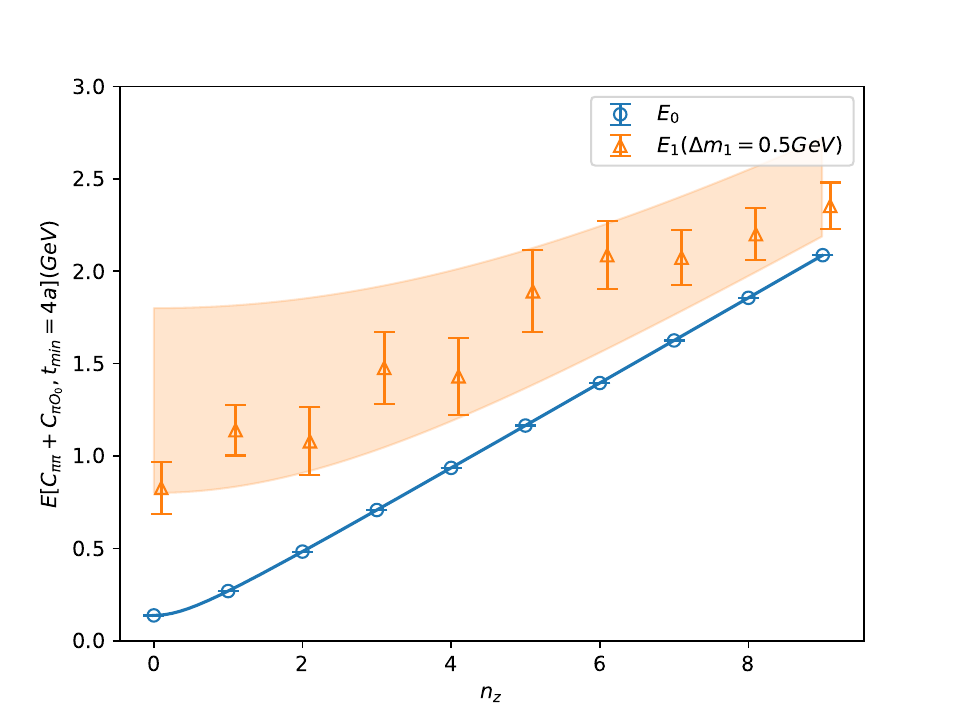}
    \caption{Dispersion relation for ground state energy $E_0$ and the first excited state energy $E_1$, fitted with priors. The blue and orange bands are the prior for $E_0$ and $E_1$. }
    \label{fig:dispersion}
\end{figure}

Once $E_0$ and the ground state coefficients $A_0^{O_0}$ and $A_0^{\pi}$ in Eq.~\eqref{eq:energy_expansion} are known, we can calculate bare $f_\pi$,
\begin{align}
    f^{\rm bare}_\pi=\frac{A_0^{O_0}(0)}{\sqrt{E_0A_0^{\pi}/2}}.
\end{align}
Although we have not calculated the renormalization constant $Z^A$ and $Z^S$ to obtain the renormalized $f_\pi$, which makes it infeasible to directly compare with the physical value, the consistency of $f^{\rm bare}_\pi$ from different fits can be used as a criteria of the fit quality besides the $\chi^2/d.o.f$ and p-value. 
The $f^{\rm bare}_\pi$ calculated from the coefficients of these different fits are shown in Fig.~\ref{fig:fpi}. The plot suggests that the fitted $f_\pi$ is consistent for different momenta with different choices of $t_{\rm min}$, although the error increases significantly for large pion momenta.
\begin{figure}[!htbp]
    \centering
    \includegraphics[width=0.45\textwidth]{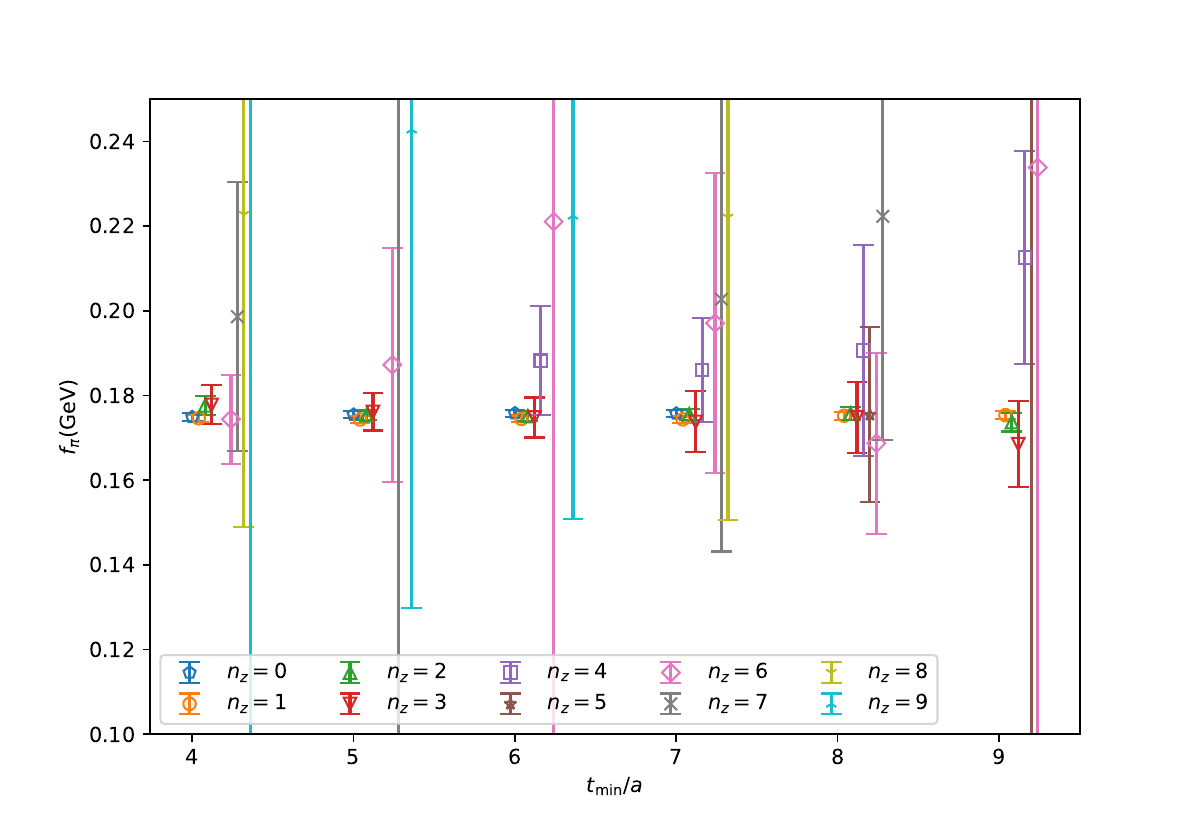}
    \caption{Bare $f_\pi$ obtained from  two-state fits for pion with different momenta as function of $t_{min}$. }
    \label{fig:fpi}
\end{figure}

Taking energies from the above two-state fit of local correlators as inputs, we fit the non-local correlations $|z|>0$ to extract the coefficients $A_0(z)$ in Eq.~\eqref{eq:kinetiC_Factors}. The bare matrix elements of $C_{\pi O_0}$ and $C_{\pi O_3}$ are both shown in Fig.~\ref{fig:bare_me}. Note that the energy spectrum are fitted separately, in order to obtain a better fit quality in each data set. 
In our fit results, both imaginary parts are consistent with zero, because the quark and anti-quark are symmetric in the pion.  Interestingly, although we can see from Eq.~\eqref{eq:kinetiC_Factors} that the coefficient $A_0^{O_3}(z)$ should statistically vanish at $P_z=0$, after normalizing the matrix elements as $H_{\gamma_z\gamma_5}(z)=A_0^{O_3}(z)/A_0^{O_3}(0)$ sample by sample, the ratio averages to non-zero values. So we are able get $P_z=0$ matrix elements for $C_{\pi O_3}$ although with large errors.

\begin{figure}[!htbp]
    \centering
    \includegraphics[width=0.45\textwidth]{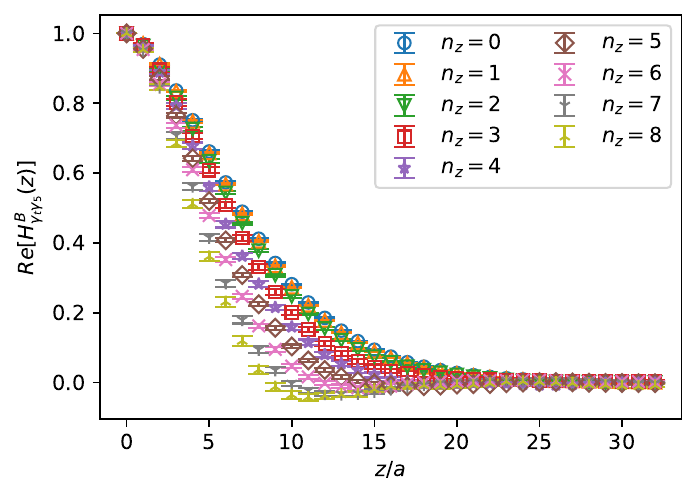}
    \includegraphics[width=0.45\textwidth]{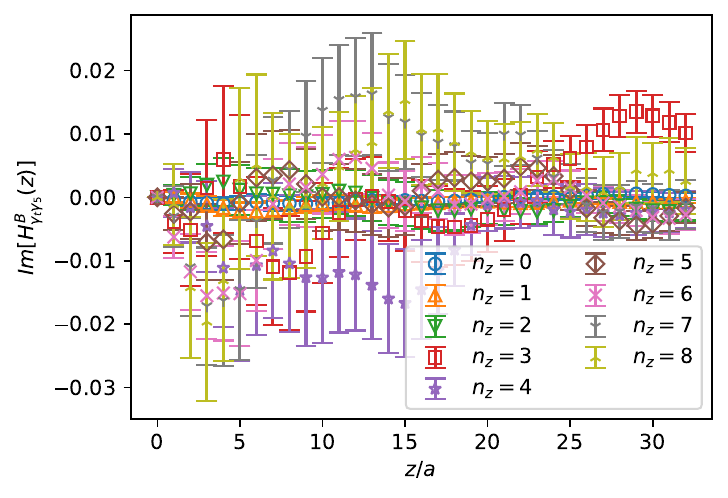}
    \includegraphics[width=0.45\textwidth]{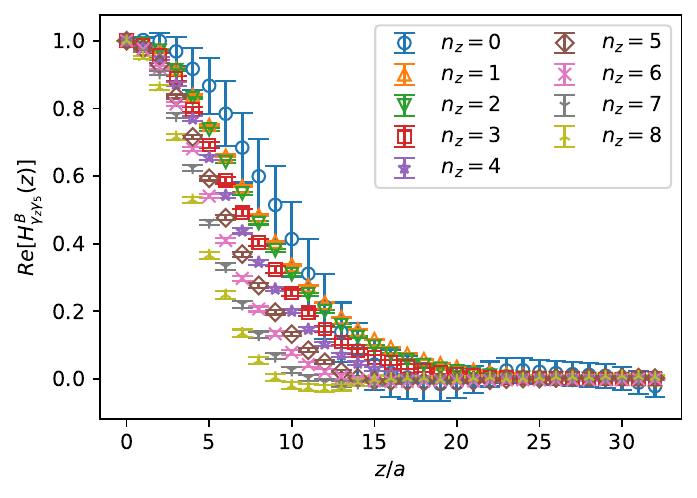}
    \includegraphics[width=0.45\textwidth]{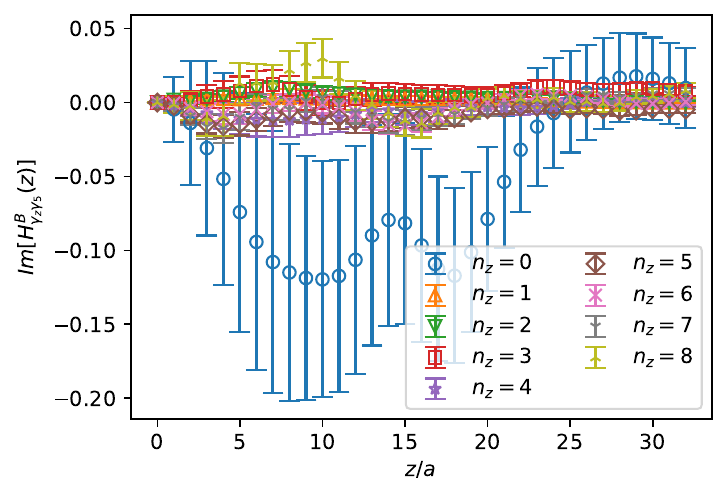}
    \caption{Matrix elements from two-state fit of $C_{\pi O_0}$ (top) and $C_{\pi O_3}$ (bottom). }
    \label{fig:bare_me}
\end{figure}

\section{Extraction of pion quasi-DA}\label{sec:quasi_da}
\subsection{Moments of pion DA}\label{sec:mom}
The quasi-DA correlator has been proven to be multiplicatively renormalizable~\cite{Ji:2017oey,Green:2017xeu,Ishikawa:2017faj},
\begin{align}
    H^R(z,P_z)=H^B(z,P_z,a)/Z(z,a)\,,
\end{align}
{where the renormalization factor $Z(z,a)$ includes the linear and logarithmic UV divergences.}
Since the UV divergences are exactly the same for the same $z$ at any $P_z$, we can cancel $Z(z,a)$ by taking the ratio of matrix elements at two different momenta{~\cite{Orginos:2017kos,Fan:2020nzz}},
\begin{align}
    \mathcal{M}(z,P_1,P_2)&=\lim_{a\to0}\frac{H^B(z,P_2,a)}{H^B(z,P_1,a)}=\frac{H^R(z,P_2)}{H^R(z,P_1)} \,.
    \label{eq:ratio}
\end{align}
{The above ratio is renormalization group invariant, so we suppress the $\mu$ dependence in the $\overline{\rm MS}$ renormalized correlation function $H^R(z,P_z)$}, which can be factorized at short-distance $z\ll \Lambda_{\rm QCD}^{-1}$ with OPE~\cite{Gao:2022vyh} 
\begin{align}\label{eq:ope_da}
    H^R(z,P_z)&=\sum_{n=0}^{\infty}\frac{(\frac{-izP_z}{2})^n}{n!}\sum_{m=0}^{n}C_{nm}(z,\mu)\langle\xi^m\rangle(\mu)+\mathcal{O}(z^2\Lambda_{\rm QCD}^2),
\end{align}
which depends on the non-perturbative moments of the light-cone-DA,
\begin{align}
    \langle\xi^m\rangle(\mu)=\int_0^1 dx (2x-1)^m\phi(x,\mu),
\end{align} 
and the perturbative matching coefficients $C_{nm}(z,\mu)$. It allows us to extract the moments $\langle\xi^m\rangle(\mu)$ from the ratio $\mathcal{M}(z,P_1,P_2)$ when the higher twist correction are suppressed $\mathcal{O}(z^2\Lambda_{\rm QCD}^2)\ll1$, 
\begin{align}
    \mathcal{M}(z,P_1,P_2)\approx \frac{\sum_{n=0}^{\infty}\sum_{m=0}^{n}\frac{(\frac{-izP_2}{2})^n}{n!}C_{nm}(z,\mu)\langle\xi^m\rangle}{\sum_n^{\infty}\sum_{m=0}^{n}\frac{(\frac{-izP_1}{2})^n}{n!}C_{nm}(z,\mu)\langle\xi^m\rangle}.
    \label{eq:ratio_ope}
\end{align}
For the symmetric pion DA, only even moments of $\xi$ are non-vanishing. We can expand Eq.~\eqref{eq:ratio_ope} to $m,n=\{0,2,4\}$ at NLO~\cite{Radyushkin:2019owq,Gao:2022vyh}:
\begin{align}
    C^{\gamma_i}_{nm}=\delta_{nm}+\frac{\alpha_sC_F}{2\pi}\left(\begin{matrix}
        \frac{3}{2}L+\frac{5}{2}+\delta_{i3}& 0 & 0\\
        -\frac{5}{12}L+\frac{3}{4}+\frac{\delta_{i3}}{6}&\frac{43}{12}L-\frac{13}{4} +\frac{\delta_{i3}}{6} & 0\\
        -\frac{2}{15}L+\frac{11}{60}+\frac{\delta_{i3}}{15}&-\frac{19}{30}L+\frac{8}{5}+\frac{\delta_{i3}}{15} &\frac{68}{15}L-\frac{1247}{180}+\frac{\delta_{i3}}{15}
    \end{matrix}\right),
    \label{eq:ope_coeff}
\end{align}
where $i=0$ or $3$, $L=\ln\frac{z^2\mu^2e^{2\gamma_E}}{4}$. The triangular matrix $C$ indicates a non-multiplicative renormalization group evolution~\cite{Lepage:1980fj}. The operator itself follows the renormalization group (RG) equation
\begin{align}
    \frac{\partial \ln O}{\partial\ln\mu^2}=\gamma_C=\frac{\partial }{\partial\ln\mu^2}\ln \sum_{n=0}^{\infty}\sum_{m=0}^{n} \frac{(\frac{-izP_z}{2})^n}{n!}C_{nm}(z,\mu)\langle\xi^m\rangle(\mu),
    \label{eq:ope_rge}
\end{align}
where $\gamma_C$ is the anomalous dimension of the quark bilinear operator that has been calculated up to 3-loop order~\cite{Braun:2020ymy}. Besides, note that 
\begin{align}
    \frac{\partial \ln C_{00}}{\partial\ln\mu^2}=\gamma_C\,,
\end{align}
which can be derived from the fact that only the $n=m=0$ term remains in the $P_z=0$ matrix element. {Moreover, Eq.~\eqref{eq:ope_rge} must be satisfied for each individual term in $n$, so there is certain cancellation of the $\mu$-dependence between the coefficients and moments.}
The 1-loop evolution can be directly read from the log terms of $C^{\gamma_i}_{00}-C^{\gamma_i}_{nm}$ in Eq.~\eqref{eq:ope_coeff}:
\begin{align}
    \label{eq:evo_moment}
\frac{d}{d\ln \mu^2}\begin{pmatrix}
        1\\
        \langle \xi^2 (\mu)\rangle\\
        \langle \xi^4 (\mu)\rangle
    \end{pmatrix}
    &=\gamma_{nm}\begin{pmatrix}
        1\\
        \langle \xi^2 (\mu)\rangle\\
        \langle \xi^4 (\mu)\rangle
    \end{pmatrix}\\
    =&-\frac{\alpha_s(\mu) C_F}{2\pi}\begin{pmatrix}
        0&0&0\\
        -\frac{5}{12}& \frac{25}{12}&0\\
        -\frac{2}{15}& -\frac{19}{30}& \frac{91}{30}
    \end{pmatrix}\cdot \begin{pmatrix}
        1\\
        \langle \xi^2 (\mu)\rangle\\
        \langle \xi^4 (\mu)\rangle
    \end{pmatrix}+\mathcal{O}(\alpha_s^2).\nonumber
\end{align}
Beyond 1-loop order, the ERBL kernel has been calculated in momentum space~\cite{Dittes:1983dy}, and $\gamma_{nn}$ is the same as those of PDF operators. The off-diagonal part of $\gamma_{nm}$ has been calculated up to  3-loop order using conformal symmetry~\cite{Braun:2017cih}. Here we quote their number and convert them into the Mellin basis, for $n_f=3$,
\begin{align}
    \gamma^{(1)}_{20}=-0.0637,\ 
    \gamma^{(1)}_{40}=-0.0232,\ 
    \gamma^{(1)}_{42}=-0.0665,
\end{align}
in units of $\alpha_s^2$.
So we can either fit the moments $\{\langle \xi^2 (\mu)\rangle,\langle \xi^4 (\mu)\rangle\}$ with NLO Wilson coefficients, or fit with RG-resummed (RGR) Wilson coefficients. In the latter case, we fit the moments at an initial scale $\langle \xi^2 (\mu_i=2e^{-\gamma_E}z^{-1})\rangle$, where the log terms in $C_{nm}(z,\mu_i)$ vanish, then evolve to $\mu=2$~GeV by solving Eq.~\eqref{eq:evo_moment}. We examine the scale variation by choosing $\mu_i=2ce^{-\gamma_E}z^{-1}$ with $c=\{\sqrt{2},1,1/\sqrt{2}\}$ to estimate the uncertainties from higher-order perturbation theory.
The fitted ratio and the extracted moments are shown in Fig.~\ref{fig:moment_fit}. In the left figure, the difference between the fitted ratios from the NLO and NLO+RGR 
Wilson coefficients are negligible, because the two different fits are eventually optimized to the same function of $(zP_z)^2$ that best describes the data. The moments are then solved from these same coefficients of $(zP_z)^{2n}$ with different Wilson coefficients, thus they could be quite different. We show the fitted moments with both statistical and scale variation error bars. Note that the scale variation becomes large when $z$ increases, because the scale $z^{-1}/\sqrt{2}\sim\Lambda_{\rm QCD}$ becomes non-perturbative. The fact that the fit results depend on the $z$-value of the data indicates a non-trivial discretization effect, and it could only be determined accurately when multiple lattice spacings are included, or on fine lattice spacings where we have more reliable data points $z\gg a$ while still stays in the perturbative region $z\ll\Lambda^{-1}_{\rm QCD}$ .
\begin{figure}[!htbp]
    \centering
    \includegraphics[width=0.45\textwidth]{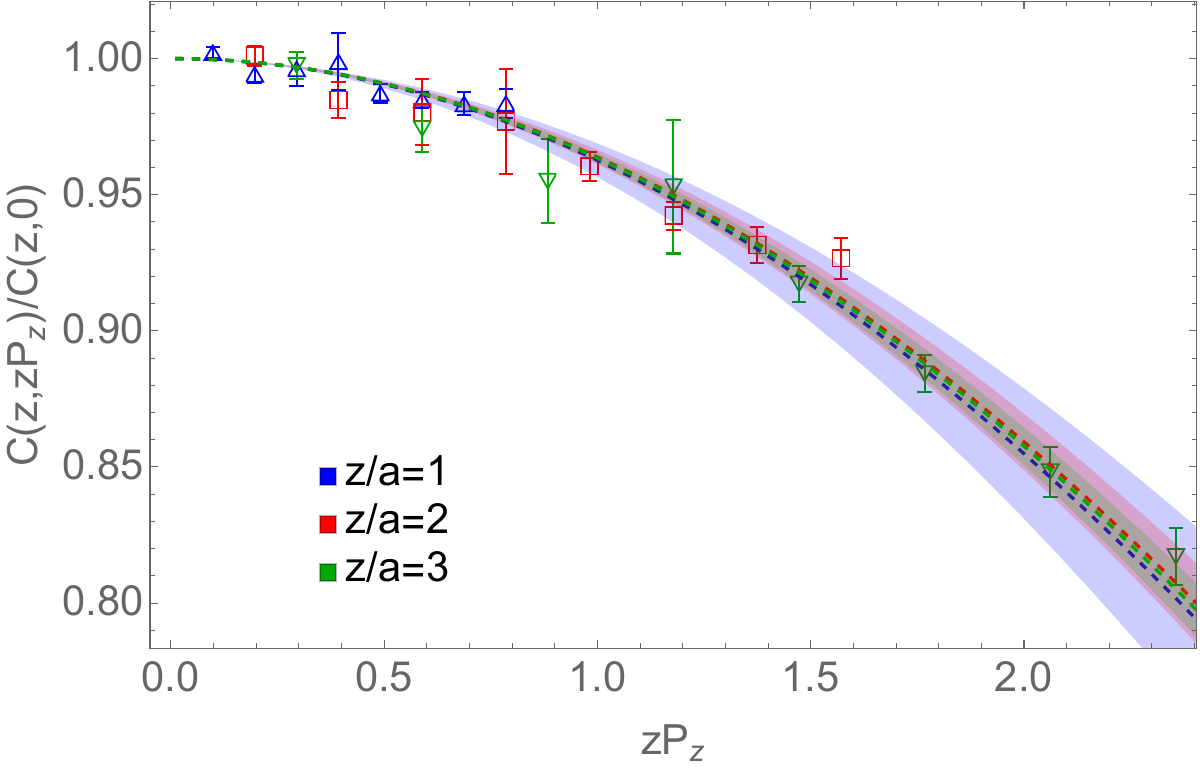}
    \includegraphics[width=0.45\textwidth]{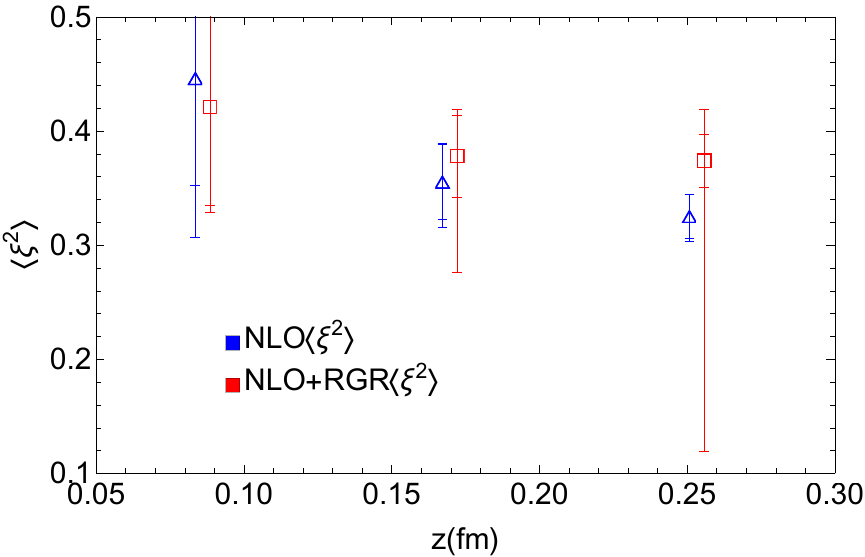}
    \caption{The fitted ratio (left) and the second moments (right) from the fits at different $z$ values. }
    \label{fig:moment_fit}
\end{figure}

\subsection{Renormalization}
In the previous section, we have extracted the moments of pion DA through a renormalization-independent approach. To directly calculate the $x$-dependence $\phi(x,\mu)$, however, we have to first properly renormalize the correlation functions.
Among multiple renormalization schemes, the hybrid renormalization scheme~\cite{Ji:2020brr} is preferred in quasi-DA calculation because it allows a perturbative matching to $\overline{\rm MS}$ scheme at all $z$, where the factorization theorem of quasi-DA is proven. In the hybrid scheme renormalization, 
\begin{align}
Z^{\rm hybrid}(z,a)=\left\{
\begin{matrix}
Z^R(z,a),\hfill  &  |z|\leq z_s \\
Z^R(z_s,a)e^{-\delta m(a) (|z|-z_s)}, & |z|>z_s
\end{matrix}
\right.
\label{eq:hybrid_renorm}
\end{align}
where $\delta m(a)\sim a^{-1}$ is the mass renormalization parameter that removes the linear divergence, and the short-distance renormalization factor $Z^R(z,a)$ can be some lattice matrix elements with the same divergence, for example, the same operator measured in other external states. A widely-used choice is the matrix element in the same hadron state at rest~\cite{Orginos:2017kos},
\begin{align}
    Z^R(z,a)=H^B(z,P_z=0,a).
\end{align}
At long range $|z|>z_s$, the hybrid scheme requires the determination of the mass renormalization parameter $\delta m(a)$. However, its determination contains an infrared ambiguity of $\sim\mathcal{O}(\Lambda_{\rm QCD})$~\cite{Ji:2020brr}, which can be labelled as a regularization scheme dependence $\delta m(a,\tau)$. When combined with the renormalon ambiguity in the perturbative matching kernel $C(x,y,\mu,P_z)$, it results in a linear correction in $1/{xP_z}$ in the final extraction of light-cone DA~\cite{Ji:2020brr}. It is proposed to absorb the linear ambiguity into a non-perturbative parameter $m_0$ independent of the hadron momentum~\cite{Ji:2020brr,LatticePartonCollaborationLPC:2021xdx,Zhang:2023bxs}, such that $m_0$ can be extracted from $P_z=0$ matrix elements, and is applied to the renormalization of $P_z>0$ data by replacing $\delta m\to\delta m(a)+m_0$ in Eq.~\eqref{eq:hybrid_renorm}. The value of $m_0$ is still ambiguous and $z$-dependent, except when both ambiguities in the extraction of $\delta m(a,\tau)$ and the perturbative matching kernel $C(x,y,\mu,P_z,\tau)$ are regularized in some specific scheme $\tau$. 
After removing the linear divergence in a specific regularization scheme, the $z$-dependence of short distance $z\ll \Lambda^{-1}_{\rm QCD}$ matrix elements should be well described by perturbation theory, up to power corrections and lattice artifacts. Since we have only one lattice spacing, we ignore the $z$-dependent discretization, and extract $m_0(\tau)$ by requiring that the $P_z=0$ renormalized matrix element matches the perturbatively calculated Wilson Coefficient $C_0(z,\mu,\tau)$ with renormalon regularized in scheme $\tau$,
\begin{align}
\label{eq:master_2}
    H^B(z,0,a)e^{(\delta m(a,\tau)+m_0(\tau)) z}=C_0(z,z^{-1},\tau)e^{\mathcal{-I}(z^{-1})}e^{\mathcal{I^{{\rm lat}}}(a^{-1})},
\end{align}
where $\mathcal{I}(\mu)=\int d\alpha \frac{\gamma(\alpha)}{\beta(\alpha)}|_{\alpha=\alpha_s(\mu)}$ is the renormalization group evolution factor that cancels the renormalization scheme dependence between the lattice scheme $H^B(z,P_z=0,a)$ and the $\overline{\rm MS}$ scheme result $C_0(z,\mu,\tau)$~\cite{Zhang:2023bxs}, in which the anomalous dimension $\gamma(\alpha)$ have been calculated to 3-loop order~\cite{Braun:2020ymy} and the beta function $\beta(\alpha)$ have been calculated to 5-loop order~\cite{vanRitbergen:1997va,Herzog:2017ohr} in $\overline{\rm MS}$ scheme. 
Here we apply the leading renormalon resummation (LRR) method as introduced in Ref.~\cite{Zhang:2023bxs} to regularize the linear ambiguity and extract $m_0(\tau)$, such that the linear power corrections are cancelled when a corresponding LRR-improved matching $C^{\rm LRR}(x,y,\mu,P_z,\tau)$ is applied. Note that we have only one single lattice spacing $a=0.0836$~fm, so with $a$ and $\tau$ fixed, $\delta m(a,\tau)$ is just a constant. Thus we redefine $\delta m\equiv\delta m(a,\tau)+m_0(\tau)$ and extract it together from $P_z=0$ data via Eq.~\eqref{eq:master_2}. Taking the ratio between two adjacent $z$'s, we get
\begin{align}   \label{eq:m0_define}
    a\delta m&=\mathcal{I}((z-a)^{-1})-\mathcal{I}(z^{-1})+\ln \frac{C_0(z,z^{-1},\tau)/C_0(z-a,(z-a)^{-1},\tau)}{H^B(z,0,a)/H^B(z-a,0,a)}.
\end{align}

The regularization scheme $\tau$ of the Wilson coefficient $C_0(z,\mu,\tau)\equiv C_0^{\rm LRR}(z,\mu)$ is defined as an LRR improved perturbation series with principal value (PV) prescription. At NLO, its form is~\cite{Zhang:2023bxs}
\begin{align}\label{eq:c0_lrr}
    C_0^{\rm LRR}(z,\mu)=&1+\frac{\alpha_s(\mu)C_F}{2\pi}\left(\frac{3}{2}\ln \frac{z^2\mu^2e^{2\gamma_E}}{4}+\frac{5}{2}+\delta_{i3}\right)-\alpha_s z\mu N_m (1+c_1)\nonumber\\ &+z\mu N_m\frac{4\pi}{\beta_0}   \int_{\rm 0, PV}^{\infty}du 
     e^{-\frac{4\pi u}{\alpha_s(\mu)\beta_0}}  \frac{1}{(1-2u)^{1+b}}\big(1+c_1(1-2u)+...\big),
\end{align}
for the operator  $O_{i}$, where $b=\beta_1/2\beta_0^2$ and $c_1=(\beta_1^2-\beta_0\beta_2)/(4b\beta^4_0)$ are from higher orders in the QCD beta function, $N_m(n_f=3)=0.575$ is the overall strength of the linear renormalon estimated from the quark pole mass correction~\cite{Pineda:2001zq,Bali:2013pla}.
To stay in the perturbative region and avoid higher twist contribution, in principle we should not go beyond $z=0.3$~fm. Also note that $z=a$ usually suffers from non-trivial discretization effects, as we will show in a later section. Thus we use $z=\{2a,3a\}$ to extract $\delta m$. For comparison, we also used fixed-order result $C^{\rm FO}_0(z,\mu)=\left.C_0^{\rm LRR}(z,\mu)\right|_{N_m=0}$, and the RGR result $C_0^{\rm RGR}=C^{\rm FO}_0(z,\mu=2z^{-1}e^{-\gamma_E})e^{-\mathcal{I}(2z^{-1}e^{-\gamma_E})}$ in Eq.~\ref{eq:m0_define} to extract $\delta m$. The scale variation is examined by vary the initial scale of RGR $\mu=2cz^{-1}e^{-\gamma_E}$ with $c=\{1/\sqrt{2},1,\sqrt{2}\}$, and we show the comparison for different extractions in Fig.~\ref{fig:m0_extraction}. The LRR method significantly improves the scale variation and convergence of perturbation theory. Also, the $\delta m$ obtained from two different operators are consistent, {suggesting} that our extracted $\delta m$ is universal for this Euclidean non-local quark-bilinear operator in a pion external state. Since the normalized matrix elements $H^B_{\gamma_z\gamma_5}(z,P_z=0)$ are very noisy, we take the result $\delta m=0.397^{+0.032}_{-0.008}$~GeV extracted from $H^B_{\gamma_t\gamma_5}(z,P_z=0)$ at NLO. Then the same $\delta m$ is used to renormalized both $H^B_{\gamma_z\gamma_5}$ and $H^B_{\gamma_t\gamma_5}$. 
\begin{figure}[!htbp]
    \centering
    \includegraphics[width=0.45\textwidth]{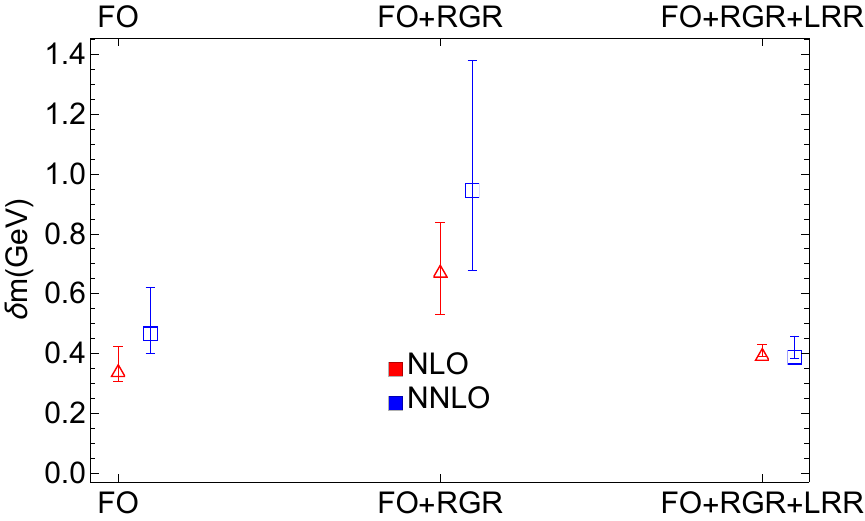}
    \includegraphics[width=0.45\textwidth]{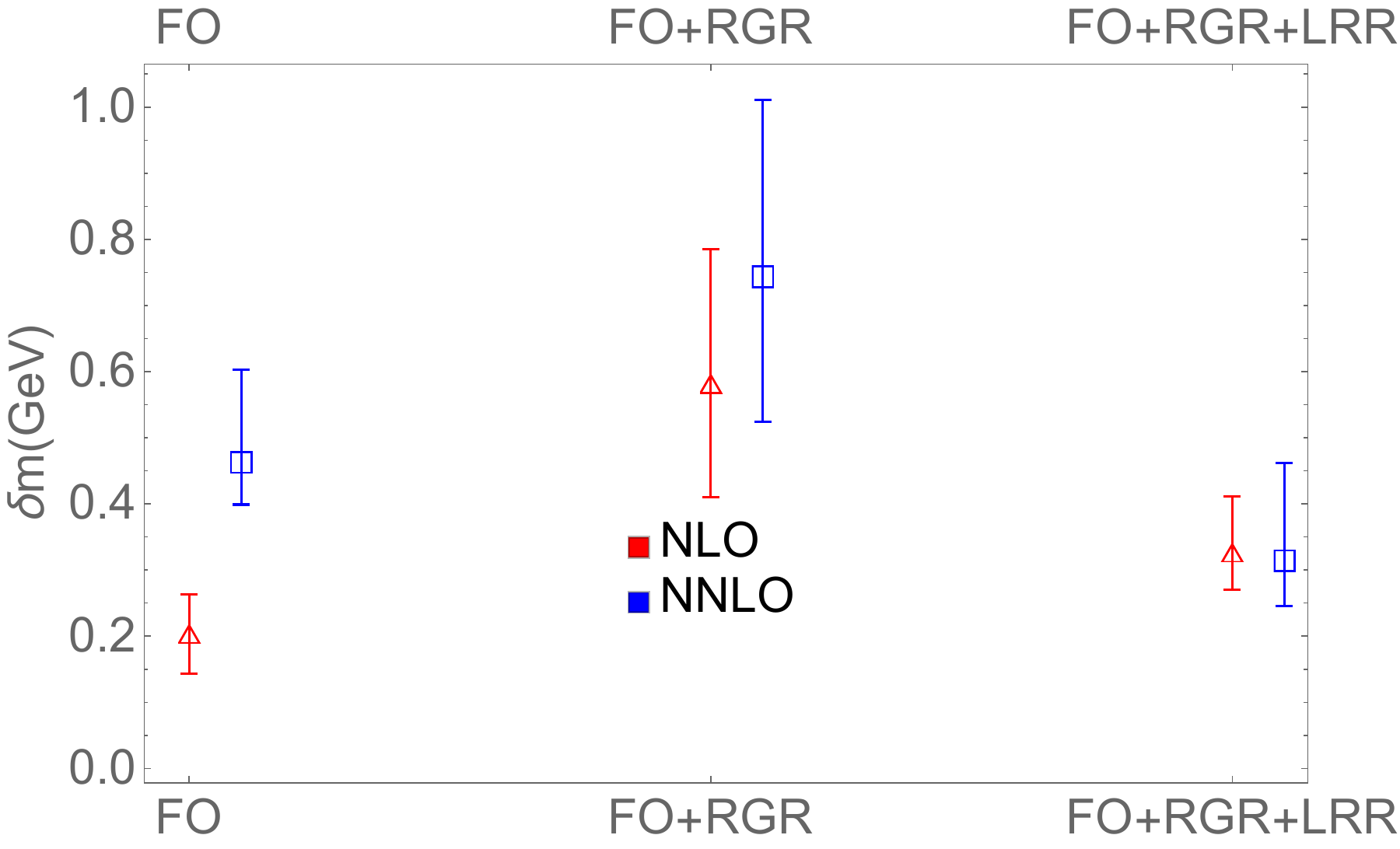}
    \caption{$\delta m$ extracted from $H^B(z,0,\gamma_5\gamma_t)$ (left) and $H^B(z,0,\gamma_5\gamma_z)$ (right) with fixed-order Wilson coefficient, RGR improved Wilson coefficient, and RGR+LRR improved Wilson coefficients. }
    \label{fig:m0_extraction}
\end{figure}

After removing the linear divergence {and regulating the renormalon ambiguity}, the matrix elements can be compared with OPE reconstruction if the first few moments are given. Taking the moment we fitted in Sec.~\ref{sec:mom}, we show the comparison of the matrix elements $e^{z(\delta m) }H^B(z,P_z,O)$ 
and RG-invariant ratio $M(z,P_0,P_z,O)$ 
with their OPE reconstruction Eq.~\eqref{eq:ope_da} and Eq.~\eqref{eq:ratio_ope}
in Fig.~\ref{fig:renorm_ope_compare}, with RGR and LRR corrections to the Wilson coefficients. We also tried to introduce higher moments in the OPE reconstruction, and found their contribution to be sub-percent level in this regime. Here we have used $P_0=0$ for $O_0$ and $P_0=0.23$~GeV for $O_3$ as the denominator when constructing the ratio. There is a good agreement for the RG-invariant ratio, but a noticeable overall deviation for renormalized matrix elements $e^{z(\delta m) }H^B(z,P_z,O_0)$ 
at $z=a$. This large deviation indicates non-negligible discretization effects at $z=a$, which appears to be universal among all momenta. When taking a ratio, such discretization effects are cancelled between two different momenta, thus the ratio is more consistent with OPE reconstruction. Nevertheless, at $z=\{2a,3a\}$, the matrix elments are still roughly consistent with OPE reconstruction, {suggesting} that the renormalization is done correctly. The consistency for both operators also suggests that $\delta m$ 
is not sensitive to the Dirac structure in the operator. Note that the discretization effects in $z=a$ matrix elements for $O_3$ are not as large  as $O_0$. So if we take a ratio of different operators,
\begin{align}
    M'(z,0,P_z,\gamma_5\gamma_z)\equiv \frac{H^B(z,P_z,\gamma_5\gamma_z)}{H^B(z,0,\gamma_5\gamma_t)},
\end{align}
extra discretization effect will be introduced by $H^B(a,0,\gamma_5\gamma_t)$. Thus when implementing hybrid scheme renormalization, $H^B(z,0,\gamma_5\gamma_t)$ is only used to renormalize $O_0$. Considering the good consistency between $H^R(z,P_z,\gamma_5\gamma_z)$ and OPE reconstruction, we assume that the discretization effect at $z=a$ is much smaller for $\gamma_5\gamma_z$, thus we can use perturbatively calculated $H^B(z,0,\gamma_5\gamma_z)$ to renormalize it, according to Eq.~\eqref{eq:master_2},
\begin{align}
    Z^R(z<z_s)=C^{\rm LRR}_{0,\gamma_5\gamma_z}(z,\mu_0)e^{-\delta m|z|+\mathcal{I}^{\rm lat}(a)-\mathcal{I}(\mu_0)},
\end{align}
where $\mu_0=2e^{\gamma_E}z^{-1}$, and $\mathcal{I}^{\rm lat}(a)$ is the evolution factor in lattice scheme, and is a constant for fixed lattice spacing that can be tuned to make sure $Z^R(z<z_s)$ is consistent with $H^B(z,0,\gamma_5\gamma_z)$. 
\begin{figure*}[!htbp]
    \centering
    \includegraphics[width=0.45\textwidth]{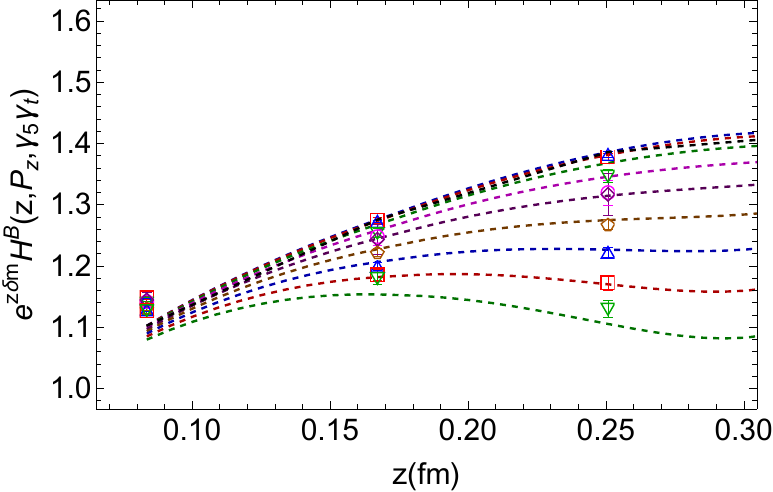}
    \includegraphics[width=0.45\textwidth]{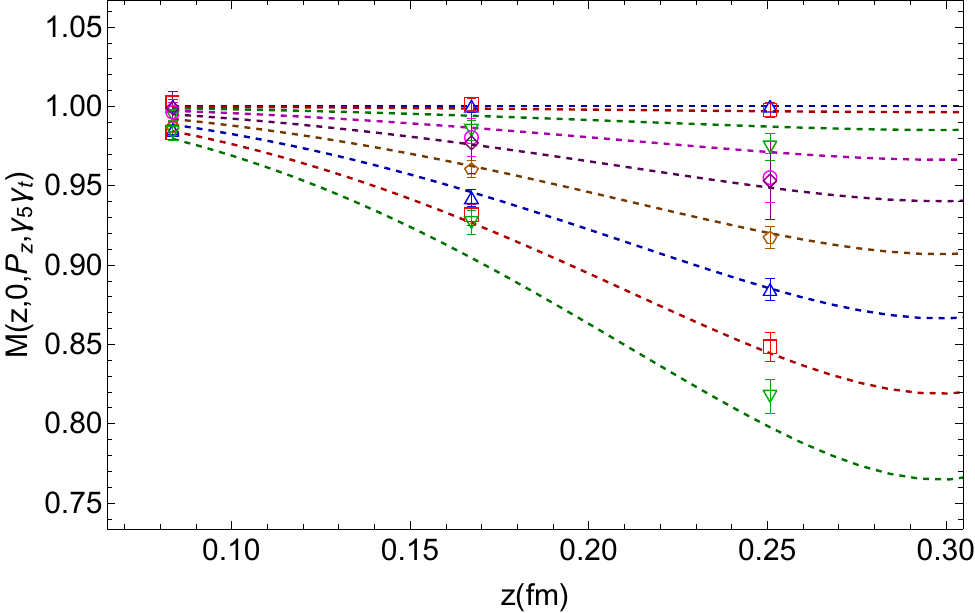}
    \includegraphics[width=0.45\textwidth]{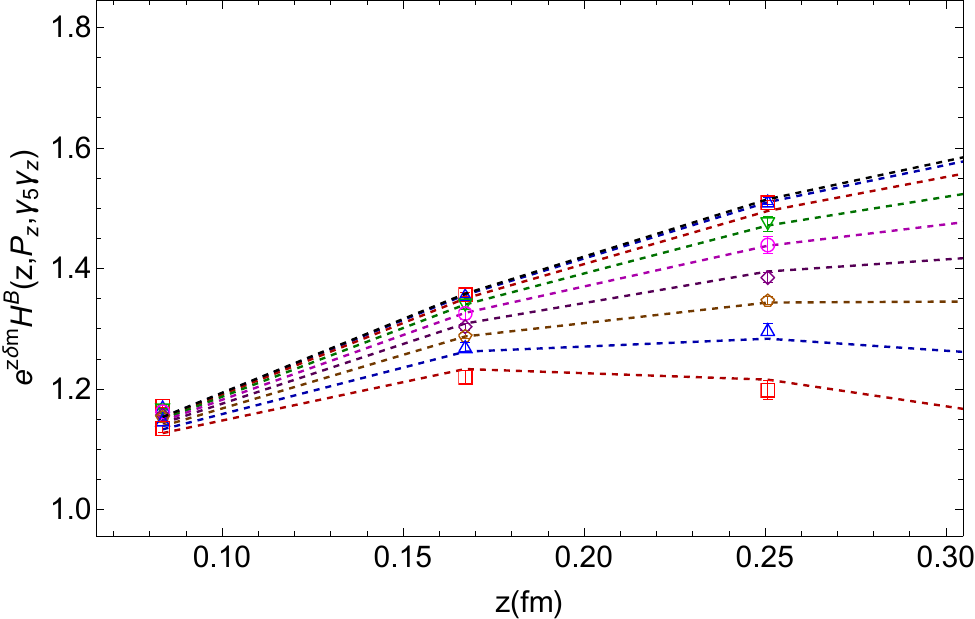}
    \includegraphics[width=0.45\textwidth]{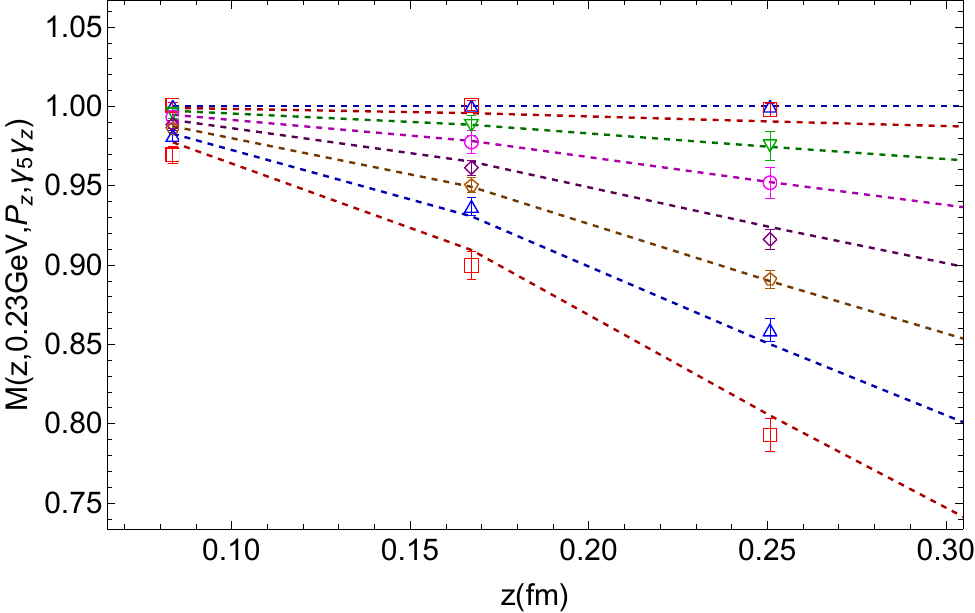}
        \caption{Renormalized matrix element (left) and the RG-invariant ratio (right) compared with OPE reconstruction in dashed lines at short distance for $O_{0}$ (top) and $O_{3}$ (bottom). Different lines correspond to different momenta. When momentum increases, the renormalized matrix elements or the ratio goes down.}
    \label{fig:renorm_ope_compare}
\end{figure*}

The renormalized matrix elements in the hybrid scheme Eq.~\eqref{eq:hybrid_renorm} are shown in Fig.~\ref{fig:renorm_me}. We find that the matrix elements at different momenta saturate to a universal shape at large $zP_z$, {except for scaling violation in $z^2$ or $1/P_z^2$ which is not distinguishable from the statistical uncertainties}.  The non-smoothness of the renormalized matrix elements near $z_s$ in Fig.~\ref{fig:renorm_me} is the nature of hybrid scheme, and has been taken into account in the perturbative matching kernel. We discuss this feature in more details in Appendix A.
\begin{figure*}[!htbp]
    \centering
    \includegraphics[width=0.45\textwidth]{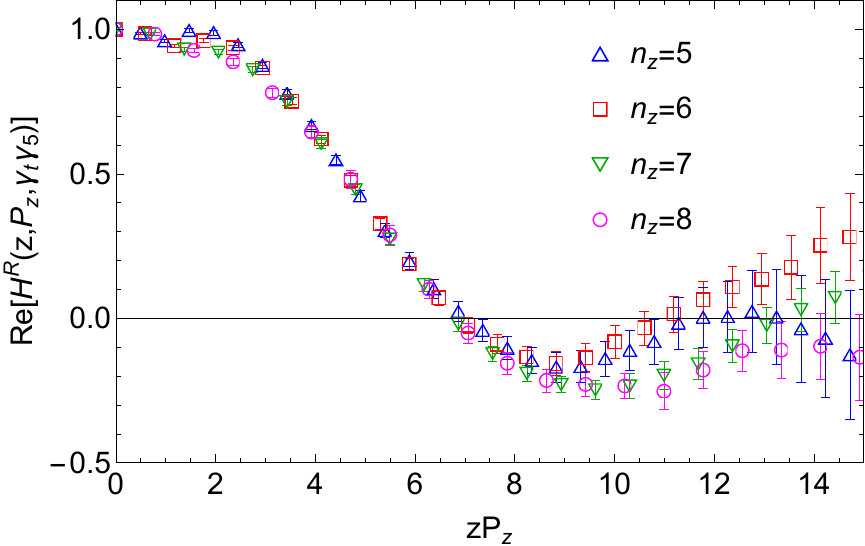}
    \includegraphics[width=0.45\textwidth]{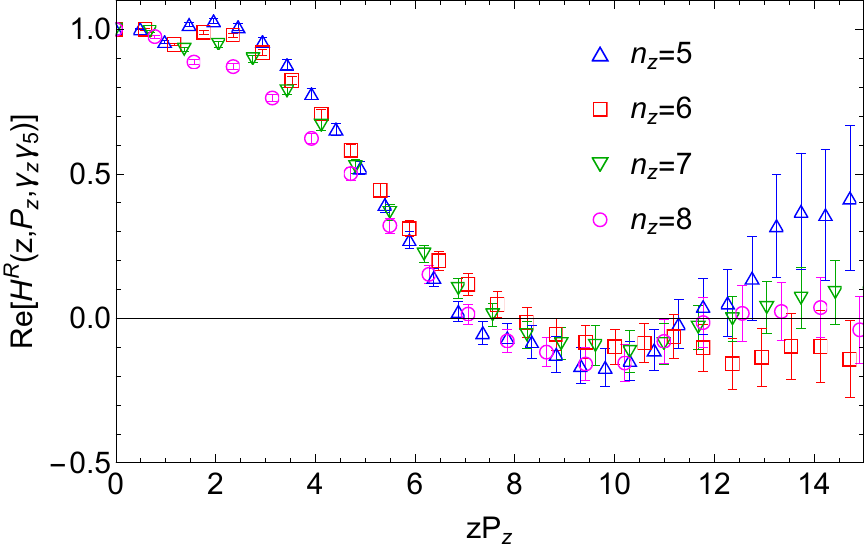}
    \caption{Real part of renormalized matrix elements for $O_0$ (left) and $O_3$ (right). At large $zP_z$, the data from different momenta follow a general curve, except for scaling violation in $z^2$ or $1/P_z^2$ which is not distinguishable from the statistical uncertainties. }
    \label{fig:renorm_me}
\end{figure*}
\subsection{$x$-dependent quasi-DA}
With the LaMET approach, we aim at calculating the local $x$-dependence {directly} in a certain range with controlled systematics. On the other hand, lattice calculations of {the matrix elements} are naturally performed in the coordinate space. Therefore, a Fourier transformation is needed to obtain the $x$-dependence of quasi-DA first. Due to the {worsening} signal-to-noise ratio {as $z$ increases}, lattice data are limited to a certain range of $z$ or $zP_z$, which makes an exact Fourier transform impossible. However, although the long tail of the correlation is unknown, it follows certain physical constraints that prevent it from going {out of control}~\cite{Ji:2020brr}. 
{By performing an extrapolation in $zP^z$ using such constraints, one can reduce the uncertainties in the long-tail region, thus eliminating unphysical oscillatory behaviors in the $x$-space after the Fourier transform}. One strong physical constraint on the long-tail distribution is the finite correlation length $\lambda_0 \propto P_z$ in the coordinate space, which results in an exponential decay $\exp{[-\lambda/\lambda_0]}$ of the Euclidean correlation functions at large $\lambda=zP_z$. 
{For example}, we can model the long-tail region to the following form~\cite{Ji:2020brr},
\begin{align}\label{eq:longtail_form}
    H(\lambda)\xrightarrow{\lambda\to\infty}\left(\frac{c_1e^{-i\lambda/2}}{(-i\lambda)^{d_1}}+\frac{c_2e^{i\lambda/2}}{(i\lambda)^{d_2}}\right)e^{-\lambda/\lambda_0},
\end{align}
{where the parameterization inside the round brackets is motivated from the Regge behavior~\cite{Regge:1959mz} of the light-cone distribution near the endpoint regions.} For symmetric pion DA, $c_1=c_2$ and $d_1=d_2$. 
To {estimate} the model dependence, we perform three different fits of the long tail, considering where the correlation functions start to be dominated by the exponentially decay, or more conservatively, assuming no exponential decay at all. These attempts include 
\begin{itemize}
    \item fitting data from $\lambda\approx 8.5$ to Eq.~\eqref{eq:longtail_form}, labeled as ``Exponential'';
    \item fitting data from $\lambda\approx 8.5$ to Eq.~\eqref{eq:longtail_form} without the exponential decay $e^{-\lambda/\lambda_0}$, labeled as ``Algebraic'';
    \item fitting data from $\lambda\approx 11$ to Eq.~\eqref{eq:longtail_form}, labeled as ``Large $\lambda$''.
\end{itemize}
The results are shown in Fig.~\ref{fig:longtail_ext}.  {The long-tail extrapolation will eventually introduce about $\pm5\%$ systematic uncertainties to the light-cone DA near $x=0.5$ in our case. To reduce this uncertainty, a more precise measurement of the long-range correlation of quasi-DA is necessary, especially for determining the oscillating behavior of the long tail.}
\begin{figure*}[!htbp]
    \centering
    \includegraphics[width=0.45\textwidth]{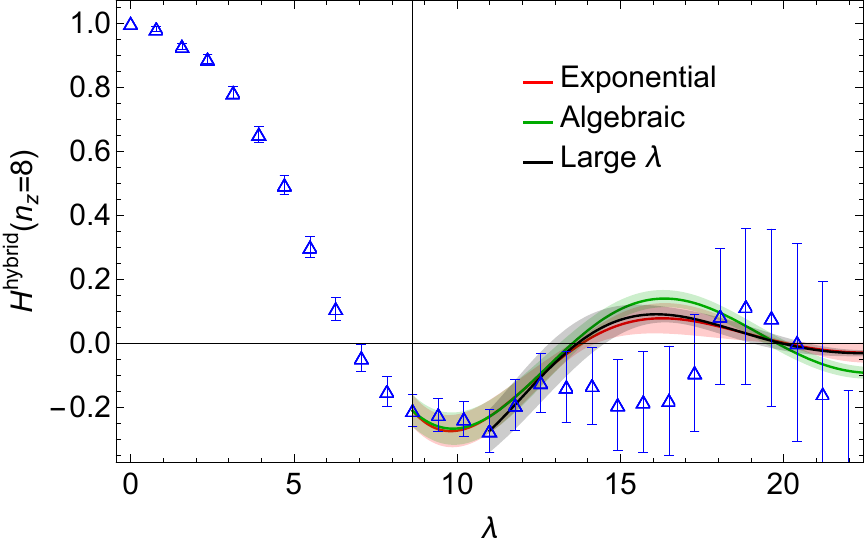}
    \includegraphics[width=0.45\textwidth]{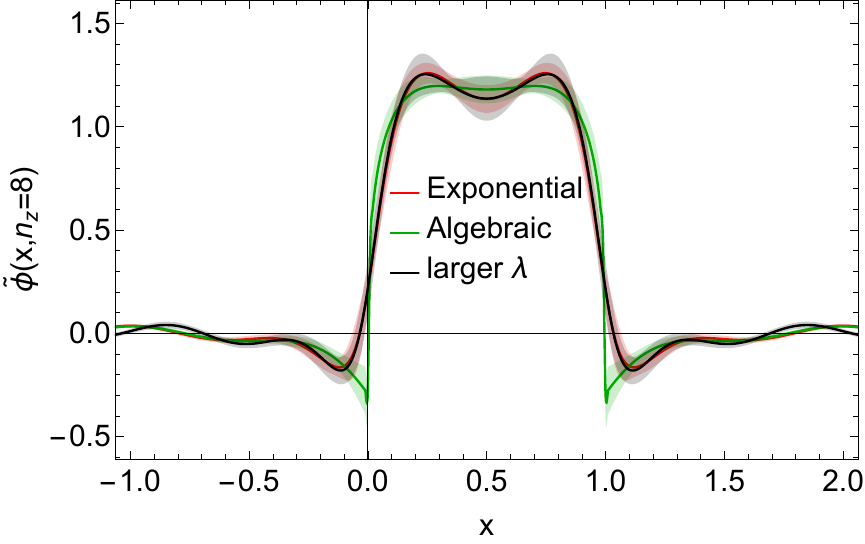}
    \caption{long-tail extrapolation and the corresponding $x$-dependent quasi-DA. The different long-tail models introduce up to $10\%$ systematic uncertainties in the $x$ dependence. }
    \label{fig:longtail_ext}
\end{figure*}

\vspace{0.2cm}
\section{Matching to the light-cone DA}\label{sec:matching}
The light-cone DA is extracted from quasi-DA through an inverse matching with power corrections starting from quadratic order in $\Lambda^2_\text{QCD}/P^2_z$ after the linear corrections are eliminated by the LRR improvement: 
\begin{align}
\label{eq:matching_mom}
 \phi(x,\mu)=&
\int_{-\infty}^{\infty} dy\ \mathcal{C}^{-1}(x,y,\mu,P_z)\tilde{\phi}(y,P_z) + \mathcal{O}\left(\frac{\Lambda^2_\text{QCD}}{x^2P^2_z},\frac{\Lambda^2_\text{QCD}}{(1-x)^2P^2_z}\right).
\end{align}
with the perturbative matching kernel $\mathcal{C}(x,y,\mu,P_z)$ calculated up to NLO~\cite{Liu:2018tox}, and in the hybrid scheme~\cite{Ji:2020brr},
\begin{align}\label{eq:matching_kernel}
    \mathcal{C}^{\gamma_t\gamma_5}(x,y,\mu,P_z)&=\delta(x-y)+\frac{\alpha_s(\mu)C_F}{2\pi}
    \left[\begin{cases}
        \frac{1+x-y}{y-x}\frac{\bar{x}}{\bar{y}}\ln\frac{(y-x)}{\bar{x}}+\frac{1+y-x}{y-x}\frac{x}{y}\ln\frac{(y-x)}{-x} & x<0\\
        \frac{1+y-x}{y-x}\frac{x}{y}\ln\frac{4x(y-x)P_z^2}{\mu^2}+\frac{1+x-y}{y-x}\left(\frac{\bar{x}}{\bar{y}}\ln\frac{y-x}{\bar{x}}-\frac{x}{y}\right) & 0< x< y<1\\
        \frac{1+x-y}{x-y}\frac{\bar{x}}{\bar{y}}\ln\frac{4\bar{x}(x-y)P_z^2}{\mu^2}+\frac{1+y-x}{x-y}\left(\frac{x}{y}\ln\frac{x-y}{x}-\frac{\bar{x}}{\bar{y}}\right) &  0<y<x< 1\\
        \frac{1+y-x}{x-y}\frac{x}{y}\ln\frac{(x-y)}{x}+\frac{1+x-y}{x-y}\frac{\bar{x}}{\bar{y}}\ln\frac{(x-y)}{-\bar{x}} & 1<x
    \end{cases}\right. \nonumber\\
    &\left.+\frac{3{\rm Si}(z_sP_z(y-x))}{\pi(y-x)}\right]^{[-\infty,\infty]}_+, 
\end{align}
where $\bar{x}=1-x$, and the plus function defined in a certain range $[a,b]$ is
\begin{align}
     [f(x,y)]^{[a,b]}_+= f(x,y)-\delta(x-y)\int_{a}^{b} f(w,y) dw,
\end{align}
and the sine integral function
\begin{align}
   {\rm Si}(x)=\int_0^x \frac{\sin y}{y}dy.
\end{align}
For $\gamma_z\gamma_5$ operator, there is an additional correction term ${\mathcal{C}}^{\gamma_z\gamma_5}={\mathcal{C}}^{\gamma_t\gamma_5}+\Delta{\mathcal{C}}^{\gamma_z\gamma_5}$,
\begin{align}
    \Delta\mathcal{C}^{\gamma_z\gamma_5}=\frac{\alpha_s(\mu)C_F}{\pi}\left[\frac{x}{y}\theta(x)\theta(y-x)+
    \begin{matrix}
        x\leftrightarrow \bar{x} \\
        y\leftrightarrow \bar{y}
    \end{matrix}\right]_+^{[0,1]}.
\end{align}
Two important systematics have been discussed in Ref.~\cite{Holligan:2023rex}, including the leading power correction and the large small-momentum logarithmic contributions. The linear power correction has been {demonstrated} to be eliminated through LRR~\cite{Zhang:2023bxs}, so we apply the same correction on the matching kernel here. On the other hand, the small-momentum logarithmic contributions can be resummed~\cite{Su:2022fiu} by solving the (ERBL) RG equation, but there is no well-established method due to the more complicated two-scale nature of the DA matching kernel~\cite{Holligan:2023rex}. 

Note that the logarithm related to quark (antiquark) momentum fraction in Eq.~\eqref{eq:matching_kernel} has the form $\frac{x}{y}\ln x$ or $\frac{\bar{x}}{\bar{y}}\ln\bar{x}$. The coefficients of the logarithm are thus suppressed when $x\to \{0,1\}$, whose contributions vanish when the quark (antiquark) momentum becomes very soft, except for the limit when $x\to y$. Therefore, the ERBL logarithm only needs to be resummed in the threshold limit $x\to y$. And as we will show below, {the resummation of small-momentum logarithms} will also become more straightforward in the threshold limit.

\subsection{Physical scales in the matching kernel: threshold limit }

It has been shown that there are two physical scales in the quasi-DA matching, corresponding to the quark momentum $2xP_z$ and antiquark momentum $2(1-x)P_z$, respectively~\cite{Holligan:2023rex}. These two scales becomes apparent when considering the threshold limit $x\to y$ of $\mathcal{C}(x,y,\mu,P_z)$, where it can be factorized into a heavy-light Sudakov form factor $H(xP_z,\bar{x}P_z,\mu)$ and a jet function $J(|x-y|P_z,\mu)$~\cite{Becher:2006mr,Ji:2023pba}, up to higher orders of $(y-x)$ (note that the leading term is singular in $(y-x)$, thus the corrections starts from the $0$-th order),
\begin{align}\label{eq:tr_factorizaion}
    \mathcal{C}(x,y,\mu,P_z)\xrightarrow{x\to y}& H(xP_z,\bar{x}P_z,\mu)\otimes J(|x-y|P_z,\mu)+\mathcal{O}((y-x)^0).
\end{align}
In coordinate space, the factorization is multiplicative. Thus the order of the Sudakov factor $H$ and the jet function $J$ in the momentum space factorization Eq.~\eqref{eq:tr_factorizaion} does not matter in the threshold limit.
The heavy-light Sudakov factor is obtained by matching the heavy-light currents to soft-collinear effective theory~\cite{Vladimirov:2020ofp}, which in our case is the product of two components from two heavy-light currents in coordinate space~\cite{Vladimirov:2020ofp,Avkhadiev:2023poz},
\begin{align}\label{eq:sudakov_factor}
    \tilde{H}_z(xP_z,\bar{x}P_z,\mu)=C_{\phi}^{-{\rm sign}(zx)}(xP_z,\mu)C_{\phi}^{{\rm sign}(z\bar{x})}(\bar{x}P_z,\mu),
\end{align}
where the signs depend on whether the momentum of the quark (antiquark) is aligned with (opposed to) the wilson line direction. 
For example,
\begin{align}\label{eq:hl_current}
    &C_{\phi}^{\pm}(p_z,\mu)=1+\frac{C_F\alpha_s(\mu)}{4\pi}\left[-\frac{1}{2}(L_z^{\pm})^2+L_z^{\pm}-2-\frac{5\pi^2}{12}\right],\qquad L_z^{\pm}=\ln\frac{4p_z^2}{\mu^2}\pm i\pi.
\end{align}
From now on we will focus on the physical region $0<x<1$ only, because the non-physical region is always far away from the threshold limit $x\to y$ when $0<y<1$.
The heavy-light Sudakov factor in the quasi-DA differs from the quasi-PDF case in two aspects: the two heavy-light vertices have different momenta in quasi-DA; their phases are opposite in the quasi-DA. Although the Sudakov factor in Eq.~\eqref{eq:sudakov_factor} explicitly depends on the momentum fraction $x$, its imaginary part has $z$-dependence through sign$(z)$ of the Wilson line direction. When Fourier transformed to the momentum space, it contributes to the matching kernel in the physical region as:
\begin{align}
    &\Im[\tilde{H}_z(xP_z,\bar{x}P_z,\mu)]={\rm sign}(z)\pi\frac{\alpha_sC_F}{4\pi}\ln\frac{x^2}{\bar{x}^2}\,,\\
    &\mathcal{F}[i\Im[\tilde{H}_z(xP_z,\bar{x}P_z,\mu)]]_{y-x}=\frac{\alpha_sC_F}{4\pi}\frac{1}{y-x}\ln\frac{x^2}{\bar{x}^2},
\end{align}
where $\mathcal{F}[i\pi{\rm sign}(z)]_{y-x}=(y-x)^{-1}$ has been used. The real part contribution is proportional to $\delta(x-y)$ when convoluted with the jet function, or equivalently, acting as an multiplicative factor to the jet function. In the threshold factorization Eq.~\eqref{eq:tr_factorizaion}, all the external-state and Dirac-structure dependence has been absorbed into the Sudakov factor. So the jet function has the same form as the quasi-PDF~\cite{Ji:2023pba}, except that the variable is $|y-x|$ instead of $|1-x/y|$ because the integral measure differs by a factor of $1/|y|$ in the two cases,
\begin{align}
    J(|x-y|P_z,\mu)=&\frac{\alpha_sC_F}{2\pi}\left[\frac{\ln\frac{4P_z^2|y-x|^2}{\mu^2}}{|y-x|}-\frac{1}{|y-x|}\ \right]^{[-1,1]}_+\nonumber\\
    &+\delta(x-y)\left[1+\frac{\alpha_sC_F}{2\pi}\left(2+\frac{\pi^2}{4}+\frac{1}{2}\ln^2\frac{4P_z^2}{\mu^2}-\ln\frac{4P_z^2}{\mu^2}\right)\right],
\end{align}
and in coordinate space,
\begin{align}
    \tilde{J}_z(z,\mu)=1+\frac{\alpha_sC_F}{2\pi}\left(\frac{1}{2}l_z^2+l_z+\frac{\pi^2}{12}+2\right),
\end{align}
where $l_z=\ln (e^{2\gamma_E}\mu^2z^2/4)$.
In the hybrid scheme, the threshold limit of the momentum-space formalism is modified by~\cite{threshold_numerical}:
\begin{align}
    \Delta \mathcal{C}&^{\gamma_i\gamma_5}_{\rm hybrid}|_{x\to y}=-\delta(x-y)\frac{\alpha_s C_F}{2\pi}\left(\frac{3}{2}\ln{\frac{z_s^2\mu^2e^{2\gamma_E}}{4}}+\frac{5+2\delta_{i3}}{2}\right).
\end{align}
We can verify that the combined contribution reproduces the ERBL logarithm as well as the finite terms of the NLO matching kernel Eq.~\eqref{eq:matching_kernel} in the threshold limit. 

The physical scales in the matching kernel become manifest after threshold factorization. The Sudakov factor incorporates the logarithms of the quark and anti-quark's momenta, and the jet function depends on the soft gluon momentum. Moreover, the quark and antiquark momentum exist in two separate heavy-light Sudakov factors $C_\phi^{\pm}(p_z,\mu)$, thus they can be resummed independently. So in the threshold limit, we are able to resum all the different logarithms in the matching kernel by solving the RG equations.

\subsection{Resummation of logarithms in the threshold limit}
The three parts in the threshold limit are resummed separately.
The Sudakov factor follows the RG equation
\begin{align}\label{eq:sudakov_evolution}
    \frac{\partial\ln C^{\pm}(p_z,\mu)}{\partial\ln\mu}=\frac{1}{2}\Gamma_{\rm cusp}(\alpha_s)L_z^{\pm}+\gamma_c(\alpha_s),
\end{align}
where $\Gamma_{\rm cusp}$ is the universal cusp anomalous dimension~\cite{Korchemsky:1987wg} known to four-loop order~\cite{Henn:2019swt,vonManteuffel:2020vjv}, here we use the three-loop result,
\begin{align}
    \Gamma_{\rm cusp}&=\frac{4\alpha}{3\pi}+\frac{\alpha^2}{27\pi^2}\left[201-9\pi^2-10n_f\right]+\frac{\alpha^3}{3240\pi^3}\left[99225 + 330 n_f \right.\nonumber\\
    &\left. - 40 n_f^2 - 12060 \pi^2 + 600 n_f \pi^2 +  594 \pi^4 + 17820 \zeta(3)- 13320 n_f \zeta(3)\right],\nonumber
\end{align}
where $\gamma_c$ is the anomalous dimension of the Sudakov factor known to 2-loop order~\cite{Ji:2021znw,Ji:2023pba,delRio:2023pse},
\begin{align}
    \gamma_c=&-\frac{2\alpha}{3\pi}+\frac{\alpha^2}{648\pi^2}\times\left[1836\zeta(3)+39\pi^2-3612+(160+18\pi^2)n_f\right]. 
\end{align}
The jet function $J(\Delta=|x-y|P_z,\mu)$ evolves as,
\begin{align}
    &\frac{\partial J(\Delta,\mu)}{\partial\ln\mu}=-\left[\Gamma_{\rm cusp}(\alpha_s)+\gamma_J(\alpha_s)\right]J(\Delta,\mu)+\Gamma_{\rm cusp}(\alpha_s)\left(\int_{ \Delta {}'< \Delta } d\Delta{}' \frac{J(\Delta{}',\mu)}{\Delta-\Delta{}'}+\int_{\Delta{}'>\Delta} d\Delta{}' \frac{J(\Delta{}',\mu)}{\Delta+\Delta{}'}\right),
\end{align}
where $\gamma_J$ is known at up to two-loop order~\cite{Ji:2023pba},
\begin{align}
    \gamma_J=&-\frac{4\alpha}{3\pi}+\frac{\alpha^2}{12\pi^2}\times\left[60\zeta(3)+\frac{23\pi^2}{3}-\frac{1396}{9}+(\frac{233}{27}-\frac{2\pi^2}{9})n_f\right].
\end{align}
From the RG equation, we can resum the Sudakov factor $ H(x,P_z,\mu)$ numerically by setting an initial scale, such as $\mu_1=2xP_z$ and $\mu_2=2\bar{x}P_z$, in $C^{\pm}(xP_z,\mu_1)$ and $C^{\mp}(\bar{x}P_z,\mu_2)$, then evolve to $\mu=2$~GeV through numerically solving the differential equation Eq.~\eqref{eq:sudakov_evolution}. 

There is also an analytical solution to the RG equation. The Sudakov factor can be decomposed into the norm and the phase,
\begin{align}
    H(xP_z,\bar{x}P_z,\mu)&=|C_{\phi}(xP_z,\mu)C_{\phi}(\bar{x}P_z,\mu)| \times e^{iA(xP_z,\mu)-iA(\bar{x}P_z,\mu)},
\end{align}
where the phase is read from the fixed-order expression of $C^{\pm}_\phi(p_z,\mu)$,
\begin{align}
    A(p_z,\mu)=\pi{\rm sign}(z)\frac{\alpha_s(\mu)C_F}{4\pi}\left(1-\ln\frac{4p_z^2}{\mu^2}\right),
\end{align}
and its evolution is just the imaginary part of Eq.~\eqref{eq:sudakov_evolution},
\begin{align}
    \frac{\partial A(p_z,\mu)}{\partial\ln\mu}=\pi{\rm sign}(z)\Gamma_{\rm cusp},
\end{align}
which can be easily solved,
\begin{align}\label{eq:sudakov_phase_rgr}
    A(p_z,\mu)=A(p_z,\mu_1)+\pi{\rm sign}(z)\int_{\mu_1}^\mu d\ln \mu'\ \Gamma_{\rm cusp}[\alpha_s(\mu')] 
\end{align}
Then the overall phase $\hat{A}$ is 
\begin{align}
    \hat{A}(xP_z,\bar{x}P_z,\mu)&=A(xP_z,\mu)-A(\bar{x}P_z,\mu)=\pi{\rm sign}(z)\frac{\alpha_s(\mu)C_F}{2\pi}\ln\frac{\bar{x}}{x}+\mathcal{O}(\alpha_s^2),
\end{align}
which turns out to be RG invariant, 
\begin{align}\label{eq:sudakov_phase_evolution}
    \frac{\partial \hat{A}(xP_z,\bar{x}P_z,\mu)}{\partial\ln\mu}&=\frac{\partial\Im[\ln C^{\pm}(\mu,xP_z)+\ln C^{\mp}(\mu,\bar{x}P_z)]}{\partial\ln\mu}\nonumber\\
    &=\pm \pi\Gamma_{\rm cusp}\mp \pi\Gamma_{\rm cusp}=0,
\end{align}
as the evolution of the imaginary parts in $C^{\pm}(\mu,p_z)$ and $C^{\mp}(\mu,p'_z)$ cancel.
So we can write the phase factor in an RG-invariant resummed form in terms of Eq.~\eqref{eq:sudakov_phase_rgr}, 
\begin{align}
    \hat{A}^{\rm RGR}(xP_z,\bar{x}P_z,\mu_1,\mu_2)=&\pi{\rm sign}(z) \left[\frac{\alpha_s(\mu_1)C_F}{2\pi}\left(1-\ln\frac{4x^2P_z^2}{\mu_1^2}\right)\right.\nonumber\\
    &\left.-\frac{\alpha_s(\mu_2)C_F}{2\pi}\left(1-\ln\frac{4\bar{x}^2P_z^2}{\mu_2^2}\right)+2\int_{\mu_1}^{\mu_2} \frac{\Gamma_{\rm cusp} }{\mu}d\mu\right],
\end{align}
where $\mu_1$ and $\mu_2$ are the physical scales in $C^{\pm}(\mu,p_z)$ and $C^{\mp}(\mu,p'_z)$, respectively. The RGE solution to the norm of the Sudakov factor is {~\cite{Stewart:2010qs}}
\begin{align}
    |H(\mu)|=&|H(\mu_1,\mu_2)|e^{S(\mu_1,\mu)+S(\mu_2,\mu)-a_c(\mu_1,\mu)-a_c(\mu_2,\mu)}\left(\frac{2xP_z }{\mu_1}\right)^{-a_\Gamma(\mu_1,\mu)}\left(\frac{2\bar{x}P_z }{\mu_2}\right)^{-a_\Gamma(\mu_2,\mu)}\,,
\end{align}
where the evolution factors are calculated from the QCD beta function and the anomalous dimensions,
\begin{align}
    &S(\mu_0,\mu)=-\int_{\alpha_s(\mu_0)}^{\alpha_s(\mu)}\frac{\Gamma_{\rm cusp}(\alpha)d\alpha}{\beta(\alpha)}\int_{\alpha_s(\mu_0)}^{\alpha}\frac{d\alpha'}{\beta(\alpha')},\nonumber\\
    &a_c(\mu_0,\mu)=-\int_{\alpha_s(\mu_0)}^{\alpha_s(\mu)}\frac{\gamma_{c}(\alpha)d\alpha}{\beta(\alpha)},\nonumber\\
    &a_{\Gamma}(\mu_0,\mu)=-\int_{\alpha_s(\mu_0)}^{\alpha_s(\mu)}\frac{\Gamma_{\rm cusp}(\alpha)d\alpha}{\beta(\alpha)}.
\end{align}

The evolution of jet function is more complicated and difficult to implement numerically directly in momentum space. In coordinate space, it is multiplicative,
\begin{align}\label{eq:jet_evolution}
    \frac{\partial\ln \tilde{J}(z,\mu)}{\partial\ln\mu}=\Gamma_{\rm cusp}(\alpha_s)l_z-\gamma_J(\alpha_s),
\end{align}
which can be used to derive an analytical solution in momentum space~\cite{Becher:2006mr}, 
\begin{align}
\label{eq:jet_fun_resum}
J&(\Delta,\mu)=e^{\left[-2S(\mu_i,\mu)+a_J(\mu_i,\mu)\right]}\tilde{J}_z(l_z=-2\partial_\eta,\alpha_s(\mu_i))\left.\left[\frac{\sin(\eta\pi/2)}{|\Delta|}\left(\frac{2|\Delta|}{\mu_i}\right)^{\eta}\right]_*\frac{\Gamma(1-\eta)e^{-\eta\gamma_E}}{\pi}\right|_{\eta=2a_{\Gamma}(\mu_i,\mu)},
\end{align}
where $\eta=2a_{\Gamma}(\mu_i,\mu)$,
\begin{align}
    &a_J(\mu_i,\mu)=-\int_{\alpha_s(\mu_i)}^{\alpha_s(\mu)}\frac{\gamma_{J}(\alpha)d\alpha}{\beta(\alpha)},
\end{align}
and the star function is defined as
\begin{align}
    \int d\Delta&\left[|\Delta|^{\eta-1}\right]_* f(\Delta) \equiv \int d\Delta |\Delta|^{\eta-1}\left(f(\Delta)-\sum_{i=0}^{\lfloor -\eta\rfloor} \frac{\Delta^i}{i!} f^{(i)}(0)\right),
\end{align}
where $\lfloor -\eta\rfloor=n$ for $n\leq -\eta < n+1$.
For $\eta>0$, which happens when $\mu_i>\mu$, the star function is trivially the function itself. For $-1<\eta<0$, which is almost always the case for $\mu_i<\mu$, the star function is the same as the plus function in the range $[-\infty,\infty]$. 

To implement the threshold resummation, we need first identify the semi-hard scale $\mu_i$ in Eq.~\eqref{eq:jet_fun_resum}. It seems natural to set $\mu_i=2|\Delta|=2|y-x|P_z$, depending on the soft gluon momentum. However, such a choice is dependent on the integrated variable $y$, and hits the Landau pole at $\Delta\to0$ for any $x$ value, thus is not numerically implementable. In analogy to the argument in deep inelastic scattering (DIS)~\cite{Becher:2006mr}, we can examine the scale by convoluting the threshold-limit kernel with a DA function and see how the logarithms in the matching kernel are converted to a logarithm depending on the light-cone-DA quark momentum fraction $x$, after the variable $y$ is integrated out. That is, to examine the following integral
\begin{align}\label{eq:da_tr_convolution}
    \tilde{\phi}(x)=\exp{\left[-2S_J(\mu_i,\mu)+a_J(\mu_i,\mu)\right]}\tilde{J}_z(\ln{\frac{\mu^2}{4P_z^2}}-2\partial_\eta,\alpha_s(\mu_i))\frac{\Gamma(1-\eta)e^{-\eta\gamma_E}}{\pi}\int_0^{1} dy \phi(y)\left[\frac{\sin(\eta\pi/2)}{|y-x|^{1-\eta}}\right]_*,
\end{align}
when $x\to 0$ and $x\to 1$. If the functional form of $\phi(y)$ is known, one could perform the integral explicitly, obtaining an analytical function of $\eta$, then figuring out the additional logarithm structure generated by the $\partial_\eta$ operator. It is indeed the case for DIS, when the integral range is limited to $y\in[x,1]$, where the PDF follows a simple power law $f(y)\propto(1-y)^b$ in the $x\to 1$ limit.
The case for DA calculation is more complicated {than the PDF case}, because the convolution of the kernel  with a DA $\phi(y)$ is always integrating over the full $y\in[0,1]$ range, as shown in Eq.~\eqref{eq:da_tr_convolution}, and it does not follow the simple power-law form in mid-$x$ region. Therefore, the same argument does not hold for DA. 
Nevertheless, we can argue that the threshold resummation effect is dominated by the integral region $y\sim x$ in Eq.~\eqref{eq:da_tr_convolution}. Firstly, the convolution of the jet function and the pion DA in Eq.~\eqref{eq:da_tr_convolution}
is mostly enhanced when $y\to x$. Meanwhile, if we perform the threshold resummation in the Mellin moment space, it is the $\ln^2 N$ and $\ln N\ln\mu$ terms that are resummed in the Wilson coefficients $C_{N\to\infty,M}(z,\mu)$. The anomalous dimensions for $C_{NM}(z,\mu)$ in the $N\to\infty$ limit have the following asymptotic behavior 
\begin{align}
    \gamma^{(0)}_{NM}\xrightarrow{N\to\infty}\left\{
\begin{matrix}
 \propto \ln N \hfill  & N=M  \\
<1 \hfill  & N>M \\
=0 \hfill  & N<M
\end{matrix}
\right.,
\end{align}
indicating that the diagonal ERBL evolution for $\langle \xi^N\rangle$ is mixed with the threshold logarithm $\ln N$, thus resulting in the effective scale $\sim Nz^{-1}$~\cite{Gao:2021hxl}, while the off-diagonal parts only contains $\ln(z\mu)$, and the physical scale is just $z^{-1}$. Therefore, the threshold resummation of quasi-DA matching will only change the diagonal Wilson coefficients $C_{NN}$, which only affects the higher moments $\langle (2x-1)^{N\to\infty}\rangle$ of the light-cone DA, determined by the distribution near the endpoint regions $x\to0$ and $x\to1$. Moreover, the contribution from the further endpoint, such as from $y\to 1$ to $x\to 0$, is suppressed compared to the $y\to0$ region, because the corresponding process requires a exchange of hard gluon.

Based on the above argument, we can limit our integration domain to be around the nearest endpoint to check the threshold resummation effect.  For example, we can integrate over $y\in[0,2x]$ for $x\to0$, where $\phi(y)$ can be modeled by  some power-law function $\phi(y)\sim y^a$,
\begin{align}
    \tilde{\phi}(x)&\xrightarrow{x\to0}e^{\left[-2S_J(\mu_i,\mu)+a_J(\mu_i,\mu)\right]}\tilde{J}_z(\ln{\frac{\mu^2}{4P_z^2}}-2\partial_\eta,\alpha_s(\mu_i))\frac{\Gamma(1-\eta)e^{-\eta\gamma_E}}{\pi}\int_0^{2x} dy y^a\left[\frac{\sin(\eta\pi/2)}{|y-x|^{1-\eta}}\right]_*\nonumber\\
    &\propto \tilde{J}_z(\ln{\frac{\mu^2}{4P_z^2}}-2\partial_\eta,\alpha_s(\mu_i))\frac{\Gamma(1-\eta)e^{-\eta\gamma_E}}{\pi} \sin(\eta\pi/2)x^{a+\eta}\nonumber\\
    &=x^{a+\eta}\tilde{J}_z(\ln{\frac{\mu^2}{4x^2P_z^2}}-2\partial_\eta,\alpha_s(\mu_i))\frac{\Gamma(1-\eta)e^{-\eta\gamma_E}}{\pi} \sin(\eta\pi/2),
\end{align}
where $\ln x$ comes from the commutator $[\partial_\eta,x^{a+\eta}]$ and is combined with $\ln (\mu^2/ (4P_z^2))$ to generate the actual physical scale $2xP_z$.
Similarly, when $x\to1$, we find
\begin{align}
    \tilde{\phi}(x) &\xrightarrow{x\to1}e^{\left[-2S_J(\mu_i,\mu)+a_J(\mu_i,\mu)\right]}\tilde{J}_z(\ln{\frac{\mu^2}{4P_z^2}}-2\partial_\eta,\alpha_s(\mu_i))\frac{\Gamma(1-\eta)e^{-\eta\gamma_E}}{\pi}\int_{2x-1}^{1} dy (1-y)^a\left[\frac{\sin(\eta\pi/2)}{|y-x|^{1-\eta}}\right]_*\nonumber\\
    &\propto \tilde{J}_z(\ln{\frac{\mu^2}{4P_z^2}}-2\partial_\eta,\alpha_s(\mu_i))\frac{\Gamma(1-\eta)e^{-\eta\gamma_E}}{\pi} \sin(\eta\pi/2)(1-x)^{a+\eta}\nonumber\\
    &=(1-x)^{a+\eta}\tilde{J}_z(\ln{\frac{\mu^2}{4(1-x)^2P_z^2}}-2\partial_\eta,\alpha_s(\mu_i))\frac{\Gamma(1-\eta)e^{-\eta\gamma_E}}{\pi} \sin(\eta\pi/2).
\end{align}
So a reasonable choice for the semi-hard scale is $2xP_z$ for $x<1/2$ and $2(1-x)P_z$ for $x>1/2$, i.e., $\mu_i=2{\rm min}(x,\bar{x})P_z$. 
To obtain the threshold-resummed matching kernel, we perform a two-step convolution: we first remove the fixed-order singular threshold terms (i.e., the Sudakov factor and jet functions in the form of $\delta (y-x)$ and $\frac{1}{y-x}$) from the fixed-order matching kernel by convoluting an inverse threshold terms with the original fixed-order matching kernel, then add back the corresponding resummed terms by convoluting with the resummed singular threshold terms,
\begin{align}\label{eq:resummed_kernel}
    C_{\rm TR}(\mu)=JH_{\rm TR}&(\mu_i,\mu_{h_1},\mu_{h_2},\mu)\otimes JH^{-1}_{\rm NLO}(\mu)\otimes C_{\rm NLO}(\mu),
\end{align}
where $JH=H\otimes J$ or $J\otimes H$, both reproducing the same singular threshold terms because the factorization is multiplicative in coordinate space.
We average them and consider the difference between the two choices as a systematic error in our final results. Here we show the example of $JH=J\otimes H$, where the momentum in $H$ labels the parton momentum in the light-cone DA. To extract the light-cone DA from quasi-DA, we need the inverse form of Eq.~\eqref{eq:resummed_kernel},
\begin{align}\label{eq:resummed_inv_kernel}
    C^{-1}_{\rm TR}(\mu)=&C^{-1}_{\rm NLO}(\mu)\otimes  J_{\rm NLO}(\mu)\otimes H_{\rm NLO}(\mu)\otimes H^{-1}_{\rm TR}(\mu_{h1},\mu_{h2},\mu)J^{-1}_{\rm TR}(\mu_i,\mu),
\end{align}
where for a specific momentum fraction $x$ in $\phi(x)$, the scales are chosen as,
\begin{align}
    \mu_i=2\min[x,\bar{x}]P_z,
    \qquad\mu_{h_1}=2xP_z,\qquad \mu_{h_2}=2\bar{x}P_z.
\end{align}
Note that we have resummed the renormalon in our perturbative matching kernel $C^{-1}_{\rm NLO}(\mu)$ with LRR to remove the linear correction. 
The same linear renormalon from the linearly divergent Wilson line self energy now exist in the jet function. To resum the large logarithms in the LRR-improved matching kernel, we then need to resum the renormalon in the jet function in exactly the same way. To implement it to the resummed form Eq.~\eqref{eq:jet_fun_resum}, it is more straightforward to modify the coordinate space expression $\tilde{J}_z$ at scale $\mu_0=2z^{-1}e^{-\gamma_E}$, then evolve to the scale $\mu$, resulting in a coefficient $(1+\frac{\alpha_sC_F}{2\pi}(\frac{1}{2}l_z^2+l_z))$ of the renormalon resummed term,
\begin{align}
    \tilde{J}&^{LRR}_z(z,\mu)=1+\frac{\alpha_sC_F}{2\pi}\left(\frac{1}{2}l_z^2+l_z+\frac{\pi^2}{12}+2\right)+\left(1+\frac{\alpha_sC_F}{2\pi}(\frac{1}{2}l_z^2+l_z)\right)\nonumber\\
    &\times N_m\Big[-\alpha_s (1+c_1)+\frac{4\pi}{\beta_0}   \int_{\rm 0, PV}^{\infty}du  e^{-\frac{4\pi u}{\alpha_s(\mu)\beta_0}}  \frac{1}{(1-2u)^{1+b}}\big(1+c_1(1-2u)+...\Big],
\end{align}
Then the regularized renormalons will cancel between the fixed-order jet function $J_{\rm NLO}(\mu_h)$ and the resummed jet function $J^{-1}_{\rm TR}(\mu_i,\mu_h)$.

To turn off the resummation of either the Sudakov factor or the jet function, we can set the corresponding scale to the same as $\mu$ in Eq.~\eqref{eq:resummed_inv_kernel}. So we can examine the effect of the two resummations independently, as shown in Fig.~\ref{fig:resum_effect}. Here we truncate the resummed results for $x\in[0.2,0.8]$ to avoid $\alpha_s(2xP_z)$ or $\alpha_s(2\bar{x}P_z)$ becoming non-perturbative.
\begin{figure}[!htbp]
    \centering
    \includegraphics[width=0.45\textwidth]{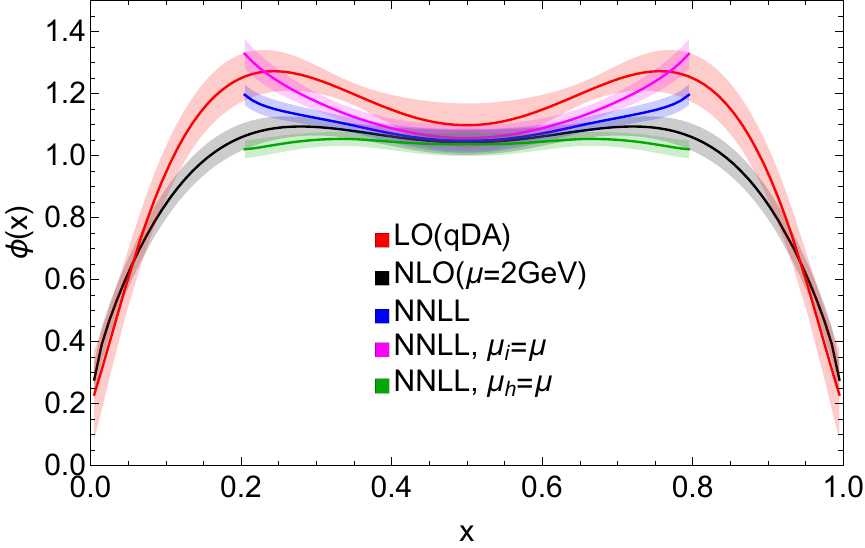}
    \caption{Matching from quasi-DA to light-cone DA with fixed order inverse kernel (black), resummed kernel (blue), and partially resummed kernels for the Sudakov factor (magenta) and the jet function (green). }
    \label{fig:resum_effect}
\end{figure}
As one can see, the resummations of the Sudakov factor and jet function have opposite effects on the final $x$-dependence of the light-cone DA. The higher-order large logarithms of soft quark (antiquark) momentum tend to enhance the endpoint regions of the DA (or equivalently, suppresses the endpoint contributions to physical observables as a convolution of hard kernel and the light-cone DA),
while the higher-order large logarithms of soft gluon momentum tend to suppress the endpoint regions. In total, the DA is enhanced at endpoints after threshold resummation when compared to a fixed-order calculation.

In order to check the convergence of perturbation theory after resummation, we vary the hard scale $\mu_h$ and semi-hard scale $\mu_i$ by a factor of $c=\{1/\sqrt{2},\sqrt{2}\}$. When the contribution from large logarithmic terms at higher orders of perturbation theory are significant, the resummed results will become sensitive to the choice of initial scales, indicating that the perturbation theory does not work or converge, so the matched DA is no longer reliable. We show the scale variation for the hard scale $\mu_h$ and semi-hard scale $\mu_i$ in Fig.~\ref{fig:scale_variation}. It is clear from the plot that the results are not very sensitive to the choice of both scales near $x=0.5$, but when approaching the endpoints, the scale dependence becomes very large, indicating that the perturbation theory breaks down. 
\begin{figure}[!htbp]
    \centering
    \includegraphics[width=0.45\textwidth]{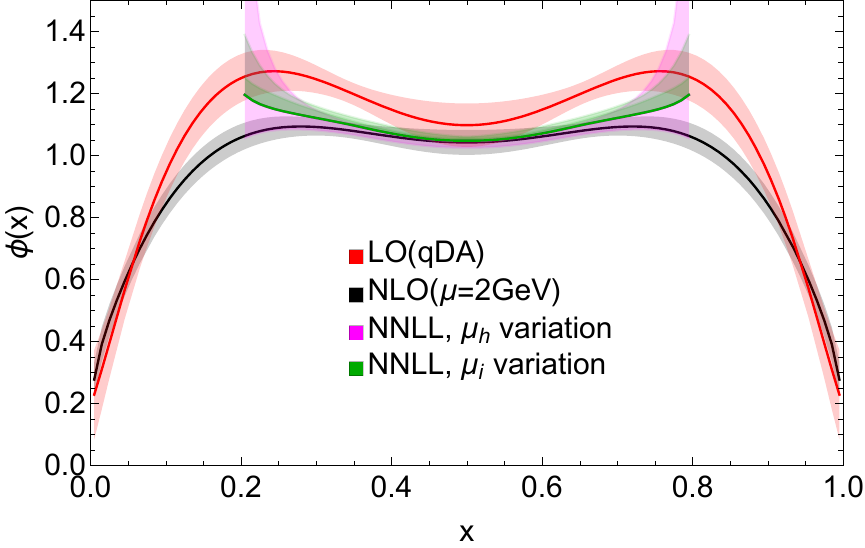}
    \caption{Scale variation for the hard scale $\mu_h$ (magenta) and semi-hard scale $\mu_i$ (green). The uncertainty is small near $x=0.5$, but becomes large when approaching the endpoint regions. }
    \label{fig:scale_variation}
\end{figure}
From our data, we have the largest pion momentum of $P_z\approx1.85$~GeV. The corresponding reliable range of prediction is estimated to be $x\in[0.25,0.75]$. To extend the range of the LaMET prediction, we have to go to higher hadron momentum. Since the range is determined by $xP_z$ and $\bar{x}P_z$, with $P_z=3.0$~GeV data we could be able to reach $x_{\rm min}\approx 0.15$.  To show that the reliable range does depend on the hadron momentum, we compare the results with $P_z\approx 1.6$~GeV in Fig.~\ref{fig:scale_variation_p7}. With a smaller momentum, the scale variation grows faster, thus reduces the reliable range of data, suggesting a similar size of $x_{\rm min}P_z$. We also notice that the results in $x\in[0.3,0.7]$ are consistent for the two momenta, suggesting that the power correction is well-controlled here. 
\begin{figure}[!htbp]
    \centering
    \includegraphics[width=0.45\textwidth]{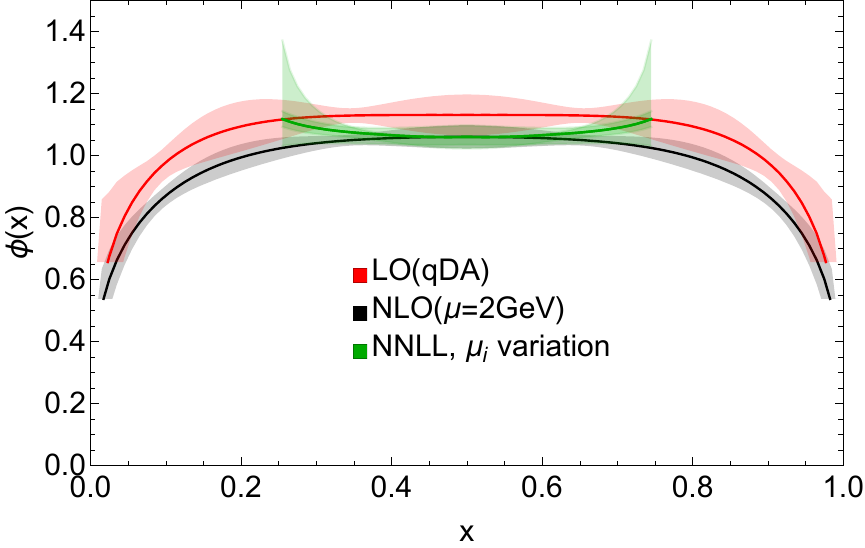}
    \includegraphics[width=0.45\textwidth]{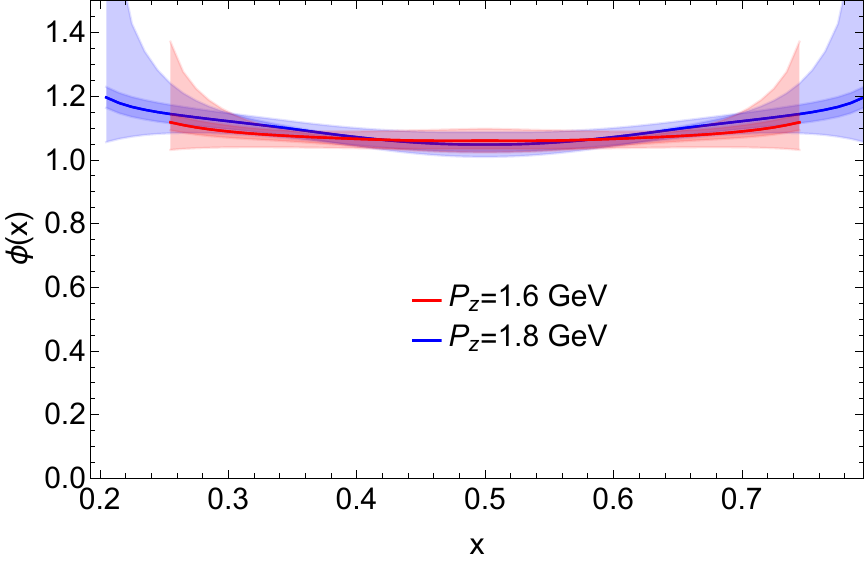}
    \caption{Left: hard scale variation for smaller pion momentum $P_z=1.6$~GeV. Right: a comparison between results from $1.8$~GeV $1.6$~GeV. }
    \label{fig:scale_variation_p7}
\end{figure}

In our final result, we consider the systematic error from the choice of $z_s\in[2a,3a]$ in the hybrid renormalization scheme, the choice of $\lambda_{\rm tail}=\{8.5,11\}$
where we start extrapolating the long tail of the coordinate space correlation functions, whether we include an exponential decaying mode in the long-tail extrapolation, and what factor $c\in\{\sqrt{2}^{-1},1,\sqrt{2}\}$ to use in the hard scale and semi-hard scale $\mu_{i,h}\to c\mu_{i,h}$. For each of these choices, we obtain $\phi_i(x)$ with a statistical error band. The systematic error band is obtained by requiring the total error band to cover all $\phi_i(x)$ results.
Figure ~\ref{fig:err_estimation} shows the relative uncertainties as a function of $x$. We find that the statistical error at $x=0.5$ is $\delta^{\rm stat}\phi(0.5)\approx 4\%$, when including the scale variation, the error increases to $\delta^{\rm stat+scale}\phi(0.5)\approx 6\%$, and the total systematic error when further including the coordinate-space extrapolation and the difference choices of $z_s$ is $\delta^{\rm stat+sys}\phi(0.5)\approx 8\%$. When approaching the endpoints, the results are dominated by the scale variations, indicating {that the perturbation theory becomes unreliable} in this region. Noticing the rapid increase of statistical uncertainty when approaching the endpoint, we truncate the result when the systematic uncertainty increases to $10\%$. This corresponding to a range of $x\in[0.25,0.75]$, within which LaMET is  able to make reliable prediction with controllable systematics.
\begin{figure}[!htbp]
    \centering
    \includegraphics[width=0.45\textwidth]{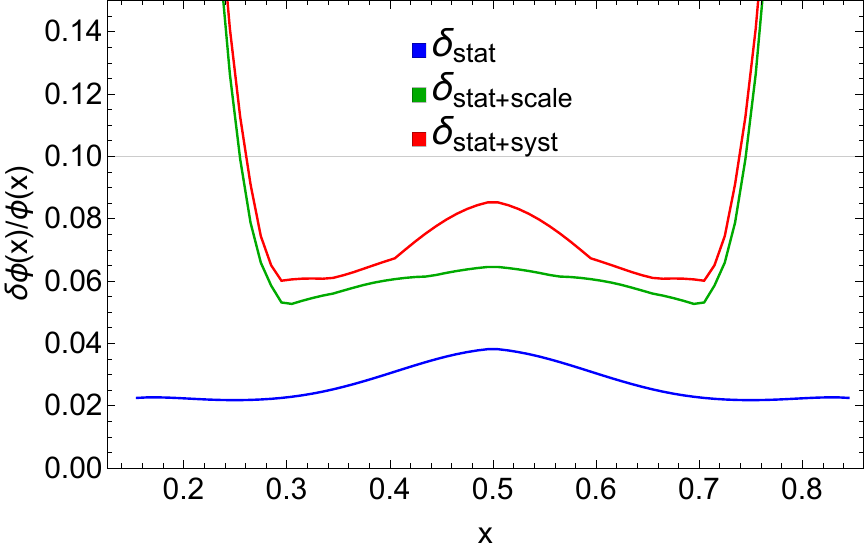}
    \caption{Relative uncertainties as a function of $x$. }
    \label{fig:err_estimation}
\end{figure}

The same procedure is applied to $O_3=\gamma_z\gamma_5$ as well. We show the two results and their comparison in Fig.~\ref{fig:final_result}. 
\begin{figure*}[!htbp]
    \centering
    \includegraphics[width=0.32\textwidth]{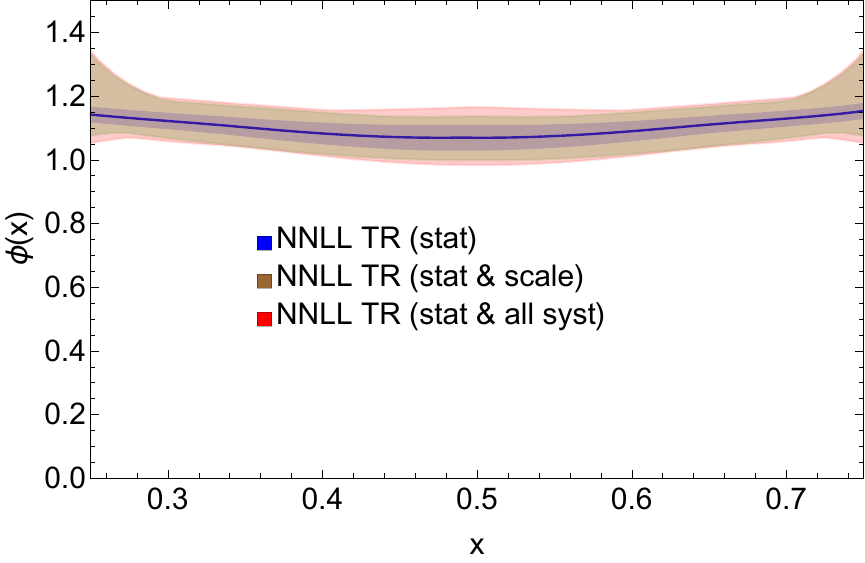}
    \includegraphics[width=0.32\textwidth]{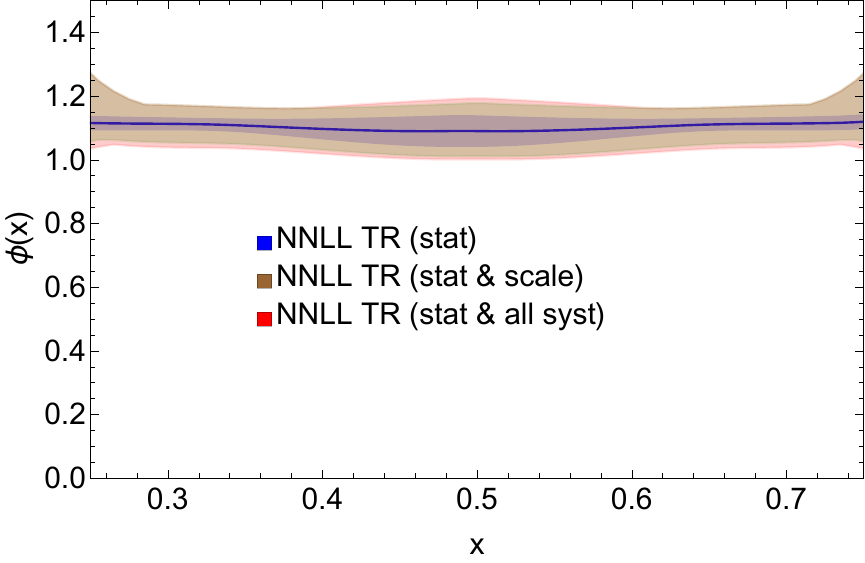}
    \includegraphics[width=0.32\textwidth]{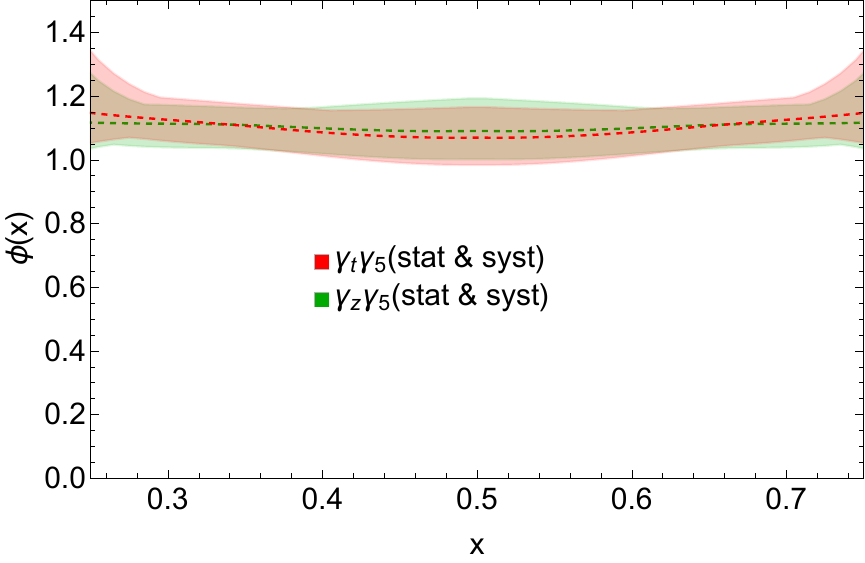}
    \caption{Final light-cone DA for $O_0$ (left) and $O_3$ (middle), including scale variation and systematic errors. And a comparison between the two operators are shown on the right. Outside the range $x\in[0.25,0.75]$, we cannot make a reliable prediction with perturbative matching.}
    \label{fig:final_result}
\end{figure*}
Both results suggest a flat DA in the mid-$x$ region.  Comparing the results of two different operators, we find they are consistent within errors. The flatness of our pion DA result with $\phi(0.5)\sim 1.1$ turns out to be qualitatively similar to the prediction from the lightcone sum rule method~\cite{Braun:1988qv,Bakulev:2001pa,Mikhailov:2021znq}.

\subsection{Comparison with data from HISQ ensembles}
~\label{sec:hisq_comparison}
We compare our final results on DWF ensembles with a similar calculation based on HISQ ensembles at physical quark masses~\cite{Gao:2022vyh,New_KaonDA} in Fig.~\ref{fig:hisq_compare}. The lattice spacing $a_{\rm HISQ}=0.076$~fm is similar to {that in} this work, and the momentum $P_z=1.78$~GeV is close to the largest one $P_z=1.85$~GeV in this calculation. The DWF results are slightly flatter when compared to the HISQ results, and the distribution at $x=0.5$ is $\phi^{\rm DWF}(0.5)=1.07(9)$, lower than the HISQ result $\phi^{\rm HISQ}(0.5)=1.19(8)$.  We also take a ratio to the central value of the DWF DA to examine the relative deviation. {The momentum-space DA on DWF ensembles is suggesting a slightly lower distribution than the HISQ ensemble within $2\sigma$ deviation. Thus, our calculation suggests that the explicit chiral symmetry breaking effect on the shape of pion DA is potential but not significant. } To further distinguish different calculations, it is important to reduce the systematic uncertainty in mid-$x$ region, which could be achieved by a more precise measurement of the long-range correlations to constrain the long-tail extrapolation. Meanwhile, this observation is based on the calculation from one lattice spacing, thus the discretization effects is not well-controlled. A more robust conclusion could be drawn after calculating on multiple lattice spacings and extrapolating the results to the continuum. 

\begin{figure}[!htbp]
    \centering
    \includegraphics[width=0.45\textwidth]{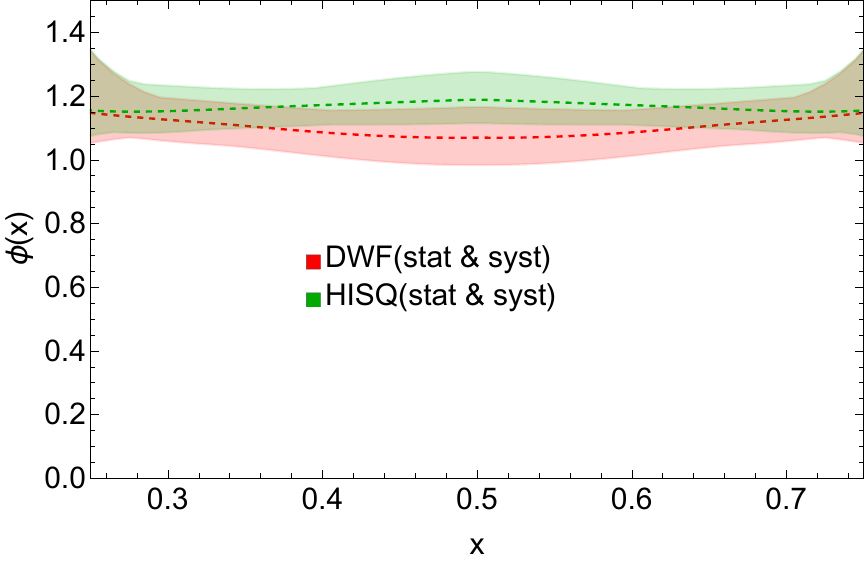}
    \includegraphics[width=0.45\textwidth]{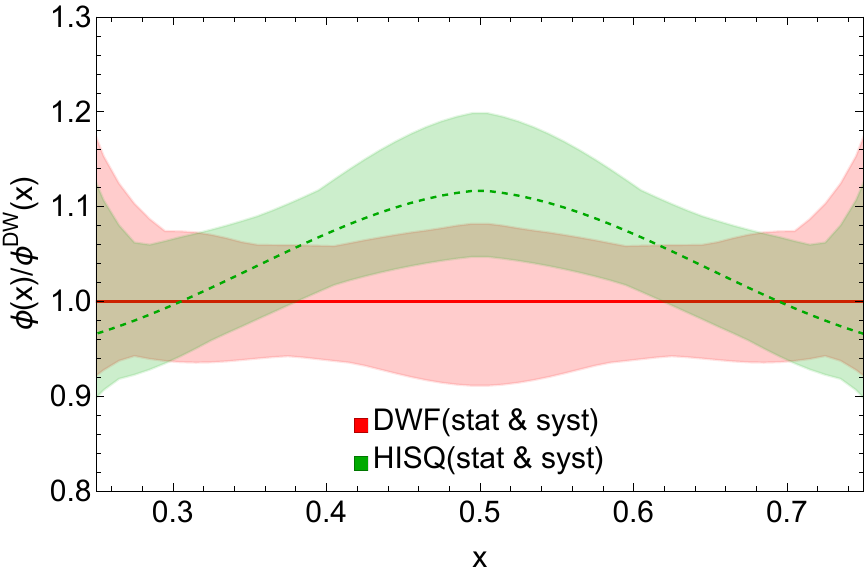}
    \caption{Comparison between the pion DAs calculated on DWF and HISQ ensembles (left) and their ratios to the central value of the DWF DA (right).}
    \label{fig:hisq_compare}
\end{figure}

\section{Conclusion}
\label{sec:conlusion}
In this article, we present the first lattice QCD calculation of the $x$-dependence of pion DA with the chirally symmetric DWF fermion action at physical pion mass. For the first time, we reformulate the threshold resummation for the quasi-DA matching kernel, and resum both the ERBL and threshold logarithms with this technique in the perturbative matching to extract the light-cone DA. We analyze two different Dirac structures $\gamma_t\gamma_5$ and $\gamma_z\gamma_5$, both of which approach the light-cone limit $\gamma_+\gamma_5$ when boosted to infinite momentum $P_z\to\infty$. With the largest pion momentum $P_z\approx1.85$~GeV, we demonstrate that the region $x\in[0.25,0.75]$ is not sensitive to scale variations, {where the systematics is under control}. Beyond this region, the scale variation becomes much larger, indicating that the perturbative matching is no longer reliable. We also show that increasing the pion momentum could extend the reliable range of $x$. Our results are consistent between the two different Dirac structures. Applying the same {analysis method} to {the quasi-DA from a} HISQ ensemble at physical point, we notice that the pion DA from the DWF ensemble is flatter, with a smaller amplitude around $x=0.5$. {The discrepancy is within $2\sigma$, thus is not statistically significant {enough to be conclusive}. }
{It should be pointed out that} our study is limited to one lattice spacing, {which cannot estimate the discretization effects} until a continuum extrapolation is performed.
Extending the future work to finer lattice spacings to examine the results in the continuum limit will give us a more convincing and comprehensive understanding of the chiral symmetry breaking effects on pion structure.

\appendix
\section{$z_s$-dependence in the hybrid renormalization scheme}
The hybrid scheme introduces a piecewise function Eq.~\eqref{eq:hybrid_renorm}, that guarantees the continuity but not smoothness of the renormalized matrix elements. Also, it introduces $z_s$-dependence to the calculation. This is not a problem in practice, because the perturbative matching kernel in Eq.~\eqref{eq:matching_kernel} also depends on $z_s$, and could cancel the $z_s$-dependence in quasi-DA, as shown in the left plot of Fig.~\ref{fig:zs_dependence}.

Meanwhile, we can also easily connect the renormalized matrix elements in hybrid schemes with different $z_s$ values through a perturbative correction, assuming $z'_s>z_s>0$:
\begin{align}
    H^{R}_{z_s}(z>z'_s,P_z,\mu)=H^{R}_{z'_s}(z>z'_s,P_z,\mu)\frac{C_0(z'_s,\mu)}{C_0(z_s,\mu)},
\end{align}
where $C_0(z,\mu)$ is the Wilson coefficient evaluated at $z$.
Then, after this perturbative correction, we find a nice agreement for $z_s=2a$ and $z_s=3a$ in coordinate space, as shown in Fig.~\ref{fig:zs_dependence}.
\begin{figure}[!htbp]
    \centering
    \includegraphics[width=0.45\textwidth]{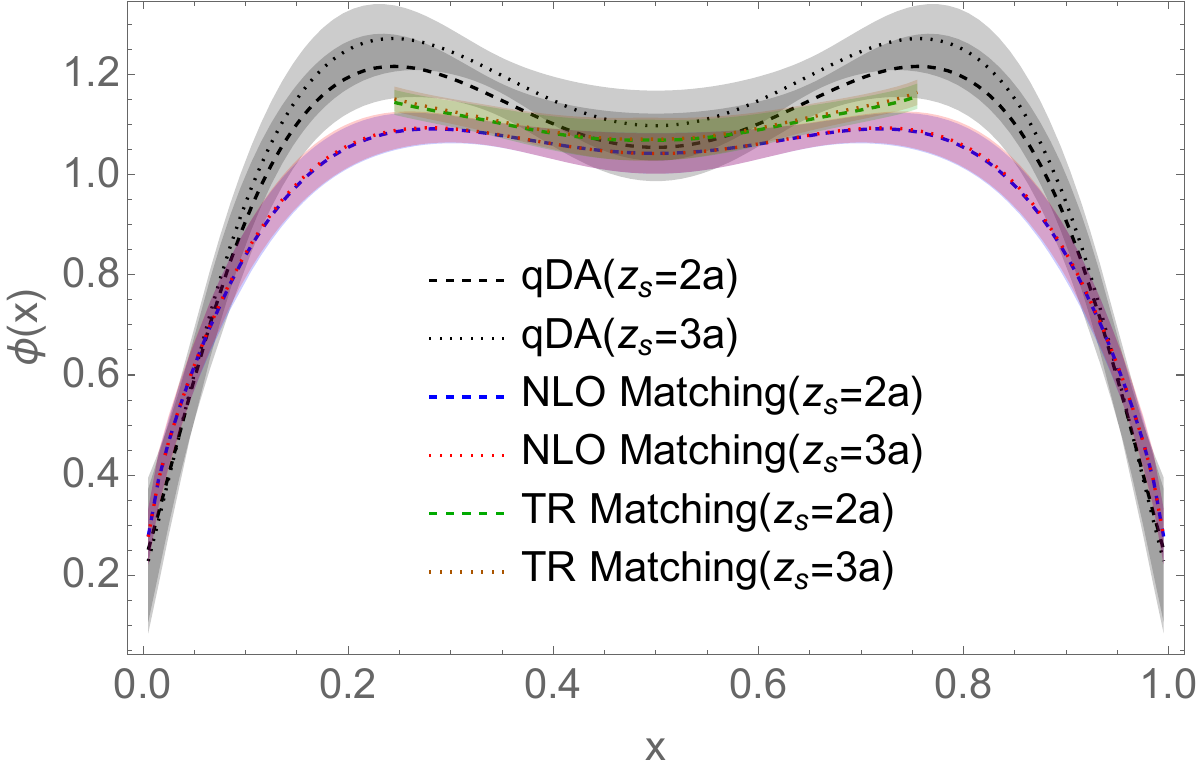}
    \includegraphics[width=0.45\textwidth]{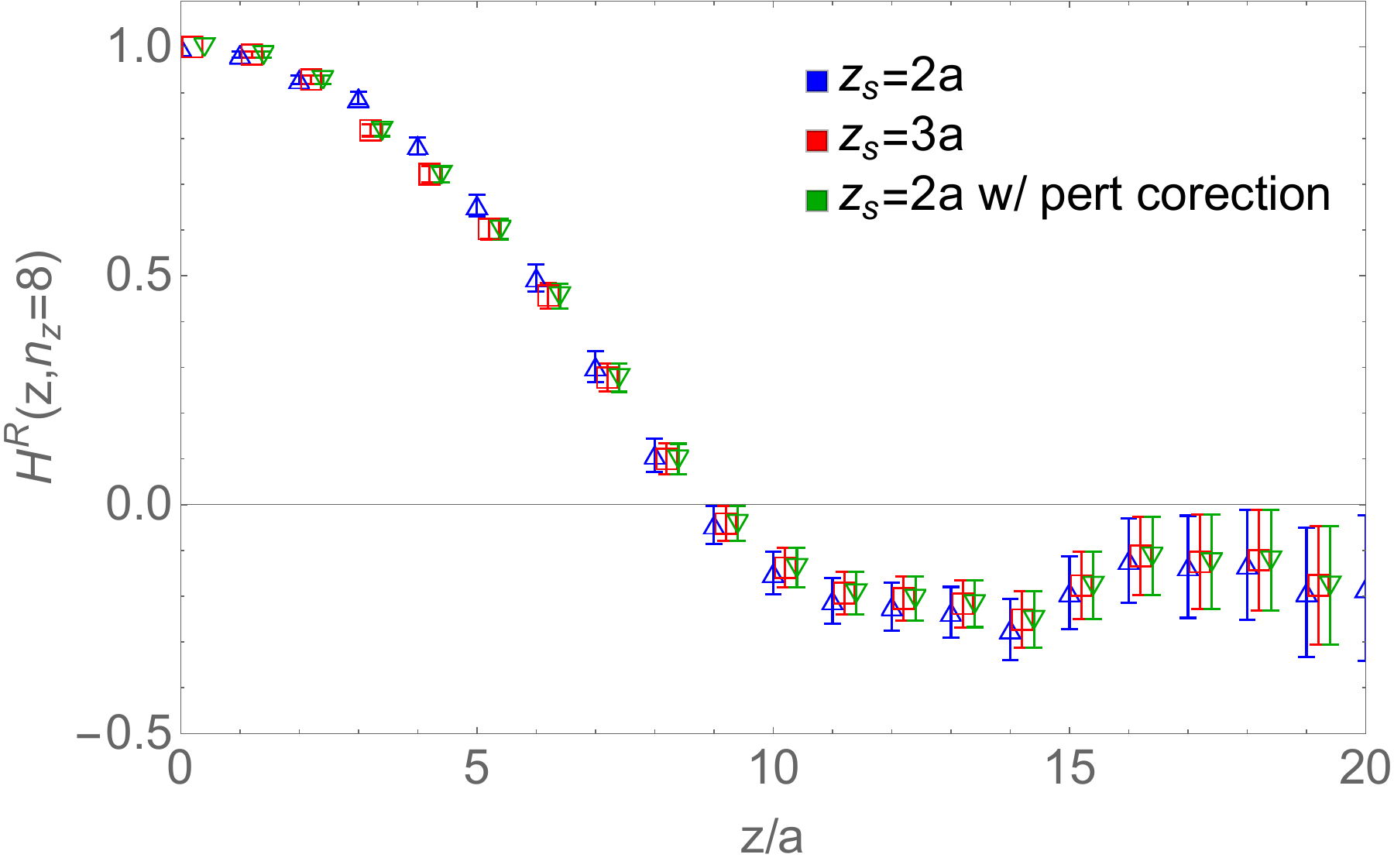}
    \caption{The $z_s$ dependence of hybrid scheme in momentum space (left) and coordinate space (right). Although the quasi-DA are different with different $z_s$ choices. The momentum-space perturbative matching or a coordinate-space perturbative correction could cancel such $z_s$ dependence. }
    \label{fig:zs_dependence}
\end{figure} 

When $z_s$ is small, the $C_0(z,\mu)$ depends logarithmically on $z_s$. Thus we would expect that the ``non-smoothness'' in the coordinate-space renormalized matrix elements comes from $\partial C_0(z,\mu)/\partial z_s$, which would be suppressed by $z$. When we choose larger $z_s$, the curve will look more smooth in Fig.~\eqref{fig:renorm_me}. But as we have just shown above, the choice of $z_s$ doesn't really affect our calculation.

\section*{Acknowledgement}
Our calculations were performed using the Grid \cite{Boyle:2016lbp,Yamaguchi:2022feu} and GPT \cite{GPT} software packages. We thank Christoph Lehner for his advice on using GPT.
We thank Yushan Su for valuable discussions.

This material is based upon work supported by The U.S. Department of Energy, Office of Science, Office of Nuclear Physics through \textit{Contract No.~DE-SC0012704}, \textit{Contract No.~DE-AC02-06CH11357}, and within the frameworks of Scientific Discovery through Advanced Computing (SciDAC) award \textit{Fundamental Nuclear Physics at the Exascale and Beyond} and the Topical Collaboration in Nuclear Theory \textit{3D quark-gluon structure of hadrons: mass, spin, and tomography}. This work was supported in part by the U.S. Department of Energy, Office of Science, Office of Workforce Development for Teachers and Scientists (WDTS) under the Science Undergraduate Laboratory Internships Program (SULI). 
YZ was partially supported by the 2023 Physical Sciences and Engineering (PSE) Early Investigator Named Award program at Argonne National Laboratory.

This research used awards of computer time provided by: The INCITE program at Argonne Leadership Computing Facility, a DOE Office of Science User Facility operated under Contract No.~DE-AC02-06CH11357; the ALCC program at the Oak Ridge Leadership Computing Facility, which is a DOE Office of Science User Facility supported under Contract DE-AC05-00OR22725; the National Energy Research
Scientific Computing Center, a DOE Office of Science User Facility
supported by the Office of Science of the U.S. Department of Energy
under Contract No. DE-AC02-05CH11231 using NERSC award
NP-ERCAP0028137.
Computations for this work were carried out in part on facilities of the USQCD
Collaboration, which are funded by the Office of Science of the
U.S. Department of Energy. Part of the data analysis are carried out on Swing, a high-performance computing cluster operated by the Laboratory Computing Resource Center at Argonne National Laboratory.

\providecommand{\href}[2]{#2}\begingroup\raggedright\endgroup

\end{document}